\documentclass[referee]{raa}            % referee version: for submission

\usepackage{graphicx,times}             %for PS/EPS graphics inclusion, new
\usepackage{natbib}
\usepackage{amssymb,amsmath}
\bibpunct{(}{)}{;}{a}{}{,}

\usepackage[a4paper=true,dvipdfm=true,pagebackref=true]{hyperref}
\hypersetup{colorlinks = true, linkcolor = green, anchorcolor = red, citecolor = blue, filecolor = red, pagecolor = red, urlcolor = red}

\begin{document}

   \title{Study of Accretion Flow Dynamics of V404 Cygni during its 2015 Outburst
%\footnotetext{$*$ Supported by the National Natural Science Foundation of China.}
}
%   \subtitle{I. Place Your Subtitle Here}

   \volnopage{Vol.0 (20xx) No.0, 000--000}      %%preserved for Editor. DOn't remove!
   \setcounter{page}{1}          %%starting page, preserved for Editor. DOn't remove!

   \author{Arghajit Jana
      \inst{1,2}
   \and Jie-Rou Shang
      \inst{3}
   \and Dipak Debnath
      \inst{2}
   \and Sandip K. Chakrabarti
      \inst{2}
   \and Debjit Chatterjee
      \inst{2}
   \and Hsiang-Kuang Chang
      \inst{4}
   }

   \institute{Astronomy \& Astrophysics Dividsion, Physical Research Laboratory, Navrangpura, 
   Ahmedabad 380009, India; {\it argha0004@gmail.com}\\
        \and
             High Energy Astrophysics, Indian Centre for Space Physics, Garia St Road, Kolkata 700084, India\\
        \and
            Institute of Astronomy, National Tsing Hua University, Hsinchu 30013, Taiwan \\
        \and
            Department of Physics, National Tsing Hua University, Hsinchu 30013, Taiwan \\
\vs\no
   {\small Received~~20xx month day; accepted~~20xx~~month day}}

\abstract{ The 2015 Outburst of V404 Cygni is an unusual one with several X-ray and radio flares
and rapid variation in the spectral and timing properties. The outburst occurred after $26$ years 
of inactivity of the black hole. We study the accretion flow properties of the source during its
initial phase of the outburst using {\it Swift}/XRT and {\it Swift}/BAT data in the energy range 
of $0.5-150$~keV. We have done spectral analysis with the two component advective flow (TCAF) model
fits file. Several flow parameters such as two types of accretion rates (Keplerian disk and sub-Keplerian
halo), shock parameters (location and compression ratio) are extracted to understand the accretion 
flow dynamics. We calculated equipartition magnetic field $B$ for the outburst and found that the 
highest $B \sim 900$~Gauss. Power density spectra (PDS) showed no break, which indicates no or very
less contribution of the Keplerian disk component, which is also seen from the result of the spectral
analysis. No signature of prominent quasi-periodic oscillations (QPOs) is observed in the PDS. This 
is due to the mismatch of the cooling timescale and infall timescale of the post-shock matter.
}

\authorrunning{Jana et al. }   
\titlerunning{Accretion Dynamics of V404 Cygni } 

   \maketitle
\section{Introduction}           %% first-level sections will be auto-capitalized
\label{sect:intro}
Transient black hole candidates (BHCs) have two phases in their lives: quiescence phase and outbursting phase.
They spend most of their lifetimes in the quiescence phase. Sudden rise in viscosity leads to an outburst when 
the X-ray intensity rises by a factor of thousands or more that of the quiescence phase. Matter from the companion 
star is accreted to the central black hole, and in this process, gravitational potential energy is converted 
to heat and radiation. Black hole (BH) spectra generally consist of two components: a multicolour blackbody bump 
and a hard power-law tail. The multicolour blackbody part is believed to originate from a Shakura-Sunyaev type 
standard thin disk \citep{SS73,NV73}. The power-law tail is believed to originate from a Compton corona 
\citep{ST80,ST85}. In the two component advective flow (TCAF) solution, the CENBOL or CENtrifugal pressure
supported BOundary Layer \citep{C95,CT95,C97} replaces the Compton corona used in other models such as disk-corona 
model \citep{Z93,HM93} or evaporated disk in ADAF \citep{Narayan1994,Esin97}. In this paper, we used the TCAF 
solution to study the accretion flow dynamics of V404~Cygni during its first outburst in 2015  after a long 
quiescent of $\sim$26 years.

V404 Cygni is one of the most studied black hole X-ray binary systems. It is also known as GS~2023+338. It was 
first identified as an optical nova in 1938 \citep{Wachmann48}. In 1956, another nova outburst was reported in 
this system \citep{Ritcher89}. In 1989, V404 Cygni went through another outburst. The 1989 outburst was discovered
with the all sky monitor onboard Ginga \citep{Makino89}. It is located at RA = $306^{\circ}.01$ and Dec = $33^{\circ}.86$.
The 1989 outburst was studied extensively. On 2015 June 15, after long 26 years in quiescent, V404 Cygni went through 
a short but violent outburst. In Dec 2015, another short activity was observed \citep{Barthelmy2015,Lipunov15}. 
The binary system V404 Cygni harbours a black hole of mass $9-12$ $M_{\odot}$ at the centre with a K-III type companion 
of mass $\sim 1$ $M_{\odot}$ \citep{Casaries1992,Shahbaz94,Khargharia10}. The inclination angle of the binary system 
is $\sim 67^{\circ}$ \citep{Shahbaz94,Khargharia10}. The orbital period of the system is $6.5$~days \citep{Casaries1992}.
The binary system is located at a distance of $2.39$~kpc, measured by parallax method \citep{Miller-Jones09}. 
V404 Cygni has a high spinning black hole with spin parameters $a^*>0.92$ \citep{Walton17}.

The 2015 outburst of V404 Cygni was discovered on June 15 simultaneously by {\it Swift}/BAT \citep{Barthelmy2015} 
and {\it MAXI}/GSC \citep{Negoro15}. During this outburst, the source was extensively observed in multi-wavelength 
bands, such as in radio \citep{Mooley15a,Trushkin15b}, optical \citep{Gazeas15} and X-ray (\cite{Rodriguez15};
\cite{Radhika16}). {\it INTEGRAL} observation reported multiple X-ray flares during the outburst \citep{Rodriguez15}. 
Several radio flares were also observed. The source showed rapid changes in the spectral properties in very short 
time \citep{Motta17}. {\it INTEGRAL} observation detected $e^{-}-e^{+}$ pair annihilation on June 20, 2015 
\citep{Siegert16,Radhika16}. {\it FERMI}/LAT detected high energy $\gamma$-ray jet in the source on June 26, 2015
\citep{Loh16}. King et al. reported detection of emission lines with Chandra-HETG, indicating strong
disc wind emission \citep{King15}

In this {\it paper}, we study the timing and the spectral properties of V404 Cygni with combined {\it Swift} XRT 
and BAT data in the broad energy range of $0.5-150$~keV during the initial phase of the 2015 outburst. We have 
done spectral analysis with the TCAF model-based {\it fits} file to extract physical flow parameters. The nature
of these model fitted accretion flow parameters allowed us to investigate physical reasons behind origin of the 
several flares, and their variability and turbulent features. We have also calculated equipartition magnetic field 
and cooling time during the outburst of the source.

The paper is organized in the following way. In \S 2, we will briefly discuss the disk structure prescribed by TCAF
and the way flow parameters decide on the spectral shape. In \S 3, we discuss the observations and the data analysis
procedure. In \S 4, we present the results of our analysis. In \S 5, we make a discussion based on our result, and
finally, in \S 6, we summarize our findings.

\section{TCAF Solution}
\label{tcaf}

TCAF configuration is based on the solution of a set of equations which govern viscous, transonic flows 
around a black hole \citep{C90}. In the TCAF solution, an accreting flow has two components: high viscous, 
high angular momentum, optically thick and geometrically thin Keplerian disk flow ($\dot{m}_d$) which 
accretes on the equatorial plane; and a weakly viscous, optically thin sub-Keplerian halo component ($\dot{m}_h$)
with low angular momentum. The Keplerian disk is immersed within the sub-Keplerian flow. Due to rise in the 
centrifugal force close to the black hole, the halo matter slows down at the centrifugal barrier and forms 
an axisymmetric shock \citep{C89}. The post-shock region or CENBOL is a `hot' and `puffed-up.' region. The 
Keplerian disk is truncated at the shock location. Multi-colour black body soft photons are generated in the
Keplerian disk. A fraction of this soft photons are intercepted by the CENBOL. Depending on the temperature
and size of the CENBOL, soft photons become hard photons via inverse-Comptonization at the CENBOL. Conversely,
some Comptonized photons reflect from the Keplerian disk and produce a reflection hump. Thus in TCAF, reflection 
component is self-consistently incorporated. However, a Gaussian line may be required to add if an iron line is 
present. CENBOL is also considered to be the base of the jets or outflows \citep{C99}. Toroidal magnetic flux 
tubes are responsible for the collimation of jet \citep{CD94,DC94}. Oscillation of CENBOL can be triggered
when the cooling and heating times inside CENBOL are similar and the emerging photons produce the quasi periodic
oscillations (QPOs) \citep{MSC96,Ryu97,C15}; hereafter C15).

\begin{figure*}
\begin{center}
\includegraphics[width=17cm,keepaspectratio=true]{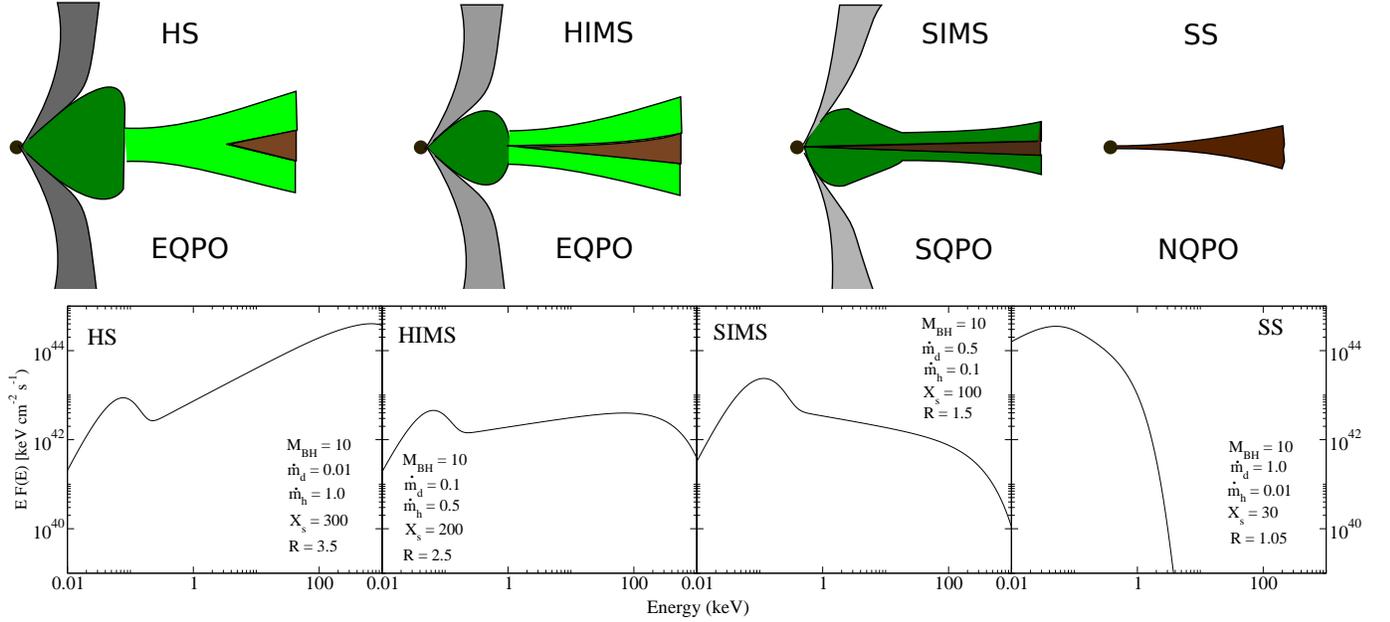}\vskip0.1cm
\includegraphics[width=18cm,keepaspectratio=true]{fig1b.eps}
\end{center}
\caption{In upper panel, cartoon diagrams of four commonly observed spectral states under TCAF paradigm are 
shown (adopted from Chakrabarti 2017). Brown, light green, dark green and grey region represent Keplerian disk, 
sub-Keplerian halo, CENBOL and jet respectively. An outbursting black hole evolves as HS $\rightarrow$ HIMS 
$\rightarrow$ SIMS $\rightarrow$ SS $\rightarrow$ SIMS $\rightarrow$ HIMS $\rightarrow$ HS. Generally both 
rising and declining HS \& HIMS show monotonically evolving type-C (low frequency) QPOs, and SIMS shows sporadic 
type-B or type-A QPOs due to shock oscillation. In soft sate these QPOs are absent, since CENBOL is absent.
Here, EQPO means evolving QPO; SQPO means sporadic QPO; NQPO means no QPO. In the bottom panel, theoretical 
spectra corresponding to the top paneled spectral states are shown. These spectra are generated using five input
parameters ($M_{BH}$ in $M_{\odot}$; $\dot{m_d}$ in $\dot{M}_{Edd}$; $\dot{m_h}$ in $\dot{M}_{Edd}$; $X_s$ in
$r_s$, $R$), whose values are marked inset.}
\label{1}
\end{figure*}

Transient BHCs generally show different spectral states during their outbursts. In TCAF, these observed
spectral states are controlled by the flow parameters (two types of accretion rates and two shock parameters).
A typical outbursting BHC generally goes through spectral state transitions to form a hysteresis loop as
follows: hard state (HS) $\rightarrow$ hard-intermediate state  (HIMS) $\rightarrow$ soft-intermediate state  
(SIMS) $\rightarrow$ soft state (SS) $\rightarrow$ soft-intermediate state  (SIMS) $\rightarrow$ hard-intermediate
state (HIMS) $\rightarrow$ hard state (HS) \citep{RM06,Nandi12,DMC15}. In the upper panel of Fig. 1 (adopted
from \citep{C18}), we show a cartoon diagram of the above four spectral states under the TCAF paradigm. In the 
lower panel, typical spectra of each spectral states correspond to the diagrams are shown. In the cartoon 
diagrams, brown, light green, dark green and grey region represent Keplerian disk, sub-Keplerian halo, CENBOL
and jet, respectively. 

Due to lower viscosity and angular momentum, the sub-Keplerian matter moves in with free-fall velocity, whereas
the Keplerian flow moves in viscous time. When an outburst is triggered, the sub-Keplerian flow dominates in the 
accretion process since it moves faster than the Keplerian disk. The Keplerian disk is truncated very far away 
by a large CENBOL. A strong shock (higher $R$) is formed at hundreds of Schwarzschild radius ($r_s$) away from 
the BH. Thus, it is difficult to cool the CENBOL by the Keplerian component. Hard X-ray flux dominates and hard 
state is observed. Compact jet is launched in this state from the CENBOL \citep{C99}. Evolving type-C QPO is 
produced in this state due to the resonance oscillation of the shock \citep{MSC96}. 

The source enters in the HIMS after HS (Fig. 1b). The Keplerian disk accretion rate continues to rise and becomes 
comparable with the sub-Keplerian halo accretion rate. As a result,  accretion rate ratio (ARR = $\dot{m}_h$/$\dot{m}_d$)
decrease. Due to rise of the Keplerian disk accretion rate, CENBOL becomes cooler, and the shock moves farther inward 
and the CENBOL shrinks. Shock strength decreases as the Compton cooling reduces the post-shock thermal pressure. 
Mass outflow rate to inflow rate ratio becomes maximum. Here also type-C QPOs are observed. 

In the SIMS (Fig. 1c), the Keplerian rate keeps on increasing, although the sub-Keplerian flow rate started to decrease.
This is because more and more sub-Keplerian flow becomes Keplerian by viscous transport. The shock becomes weak in
this state. The shock further moves in, and the CENBOL becomes small. The soft X-ray flux increases and the hard 
X-ray flux decreases in this state due to rapid rise in $\dot{m}_d$ and slow decrease in $\dot{m}_h$. Generally, 
type-A or type-B QPOs are observed sporadically in this state due to weak oscillation of the CENBOL (type-B) or 
due to oscillation of shock-less centrifugal barrier (type-A). On the SIMS to SS transition day, one may see peak 
of the Keplerian disk rate. The time (day) difference between $\dot{m}_h$ and $\dot{m}_d$ peaks gives us rough 
estimation of the viscous time scale of the source (Jana et al. 2016). In the SS (Fig. 1d), the Keplerian disk 
dominates and completely cools down the CENBOL. Soft X-ray flux dominates over hard X-ray flux. No shock is formed.
As a result, the jet is completely quenched in this state (Chakrabarti 1999, Garain et al. 2012). No QPO is produced in
the soft state. 

The flow parameters evolve oppositely during the declining phase of the outburst. Starting from SS to SIMS
transition day, both the Keplerian disk accretion rate and the sub-Keplerian halo accretion rate decreases,
although, the Keplerian disk rate decreases faster. As a result, ARR increases. As in the SIMS of the rising phase,
one may see sporadic type-B or `A' QPOs in the declining SIMS. In the declining HIMS and HS, evolving type-C QPOs
could be seen. Similar to the rising phase, one could observe compact jet in the HIMS and HS in the declining phase.

TCAF solution is implemented in {\tt XSPEC} \citep{Arnaud96} to analyze spectral properties around the black holes
\citep{DCM14,DMC15} as an additive table model. TCAF model has input parameters ($M_{BH}$, $\dot{m}_d$, $\dot{m}_h$,
$X_s$, $R$). Accretion flow dynamics around several black holes are studied quite successfully using TCAF model 
\citep{DMCM15,D17,D20,Jana16,Jana20b,C16,C19,C20,Bhattacharjee et al. 2017,Shang19}. Frequencies of the dominating
QPOs are predicted from TCAF model fitted shock parameters \citep{C16}. Masses of the black holes are estimated 
quite successfully from spectral analysis with the TCAF model \citep{Molla16,C16}. Jet contribution in the X-rays 
are also calculated using TCAF solution \citep{JCD17,Jana20a,C19}.

\section{Observation and Data Analysis}
\label{Obs}

We analyzed {\it Swift} data for 19 observations between 2015 June 15 (MJD=57188.77) and June 26 (MJD=57199.52).
We studied the source in $0.5-150$~keV energy band with combined XRT and BAT data for five observations 
(MJD = 57191.01, 57194.54, 57197.21, 57197.33, and 57198.02). The $15-150$~keV BAT data was used for four 
observations (MJD = 57188.77, 57193.56, 57198.15 and 57199.52) when only BAT observations were available. 
For the rest of the ten observations, we studied $0.5-10.0$~keV using XRT data. 

We used WT mode data for XRT observation. Cleaned event files were generated for XRT using {\tt xrtpipeline} 
command. To reduce pileup effects, we used grade-0 data. For pileup correction, we chose an annular region
around the source. We chose an outer radius of 30 pixels and a varying inner region, depending on the count 
rate. A background region is chosen far away from the source with 30 pixels radius. Then, we obtained $.pha$
and $background$ files using these cleaned event files in {\tt XSELECT v2.4}. A scaling factor was applied
to the source and background with {\tt BACKSCAL}. Spectral data were re-binned to 20 counts per bin using 
{\tt grppha} command. $0.5-10$~keV $0.01$~sec lightcurves were generated in {\tt XSELECT v2.4} using cleaned
source and background event files. We followed standard procedures to generate BAT spectra and lightcurves.
Detector plane images (dpi) were generated using the task {\tt batbinevt}. For appropriate detector quality,
we used {\tt batdetmask} task. Noisy detectors were found, and a quality map was obtained using {\tt bathotpix}. 
Then {\tt batmaskwtevt} was run to apply mask weighting to the event mode data. A systematic error was applied 
to the BAT spectra using {\tt batphasyserr}. Ray-tracing was corrected using {\tt batupdatephakw} task. Then a
response matrix for the spectral file was generated using {\tt batdetmask}. BAT lightcurves of $0.01$~sec were
obtained using {\tt batbinevt} for $15-150$~keV.

Here, we used the TCAF model-based {\it fits} file for the spectral analysis. We also used combined `diskbb' 
(DBB) and `powerlaw' (PL) models to get rough estimation about the thermal and the non-thermal fluxes where 
a reflection component is often required to find the best fits. TCAF model-based fits do not require any
additional component for reflection since the reflection component is already incorporated while generating
a spectrum. We required a $Gaussian$ model to incorporate $Fe$-$k_{\alpha}$ emission line. We used $phabs$ 
model for interstellar absorption and $pcfabs$ model for partial absorption. With the TCAF, we extracted
physical parameters such as the mass of the black hole ($M_{BH}$) in solar mass ($M_{\odot}$), Keplerian 
disk rate ($\dot{m_d}$) in Eddington rate ($\dot{M}_{Edd}$), sub-Keplerian halo rate ($\dot{m_d}$) in 
Eddington rate ($\dot{M}_{Edd}$), shock location ($X_s$) (i.e., size of the Compton cloud) in Schwarzschild
radius ($r_s$) and the shock compression ratio ($R=\rho_{+} / \rho_{-}$ with $\rho_{+}$ and $\rho_{-}$ are
post- and pre-shock density respectively). In TCAF, $N$ depends on the distance and inclination angle
of the source and is just a constant factor between the emitted flux and observed flux by a given instrument.
However, it can vary if any physical processes are present other than the accretion. Since the current version 
of TCAF model fits file does not include jets, for instance, a variation of normalization is observed if they
are present \citep{JCD17,C19}. We first analyzed the spectra after keeping the mass of the black hole as a 
free parameter. We obtained the mass of the black hole in the range of $9.5-11.5$~$M_{\odot}$ or average 
value of $10.6$ $M_{\odot}$. This measured mass range agrees very well with previously reported values by 
many authors \citep{Casaries1992,Khargharia10}. Then, we refitted all the spectra after keeping the mass of 
the black hole frozen at $10.6$ $M_{\odot}$. The result based on the later analysis is presented here.

We achieved best-fittings using {\tt steppar} command. After obtaining a best-fit based on $\chi^2_{red}
(\sim 1)$ with TCAF, we ran {\tt steppar} to verify fitted parameter values. The `steppar' command ran for 
pair of parameters $\dot{m}_d$-$\dot{m}_h$ and $X_s$-$R$. We also calculated uncertainties with the {\tt steppar}. 
In Figs. 2 and 3, 2D-contour plots for two observations (MJD=57188.77 \& 57189.62) are shown. They are quite
satisfactory.

\begin{figure}
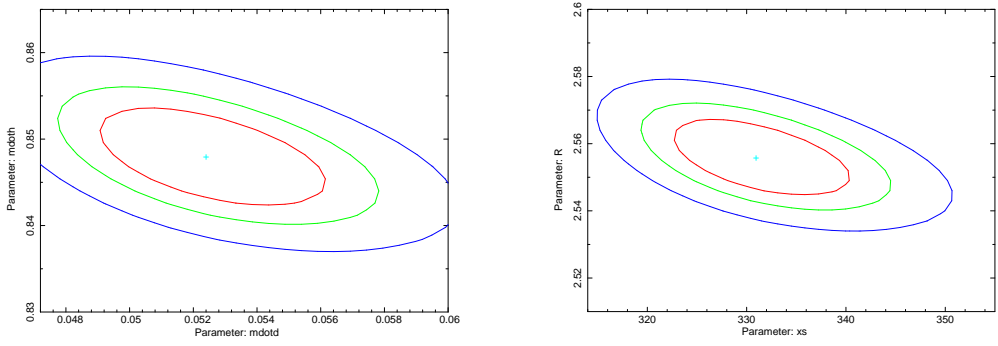

 \hbox{
       \hspace*{0.2cm}\includegraphics[angle=270,width=7cm]{md-mh-cntr-bat.ps}
       \hspace*{0.2cm}\includegraphics[angle=270,width=7cm]{xs-r-cntr-bat.ps}
        }
 \caption{2D confidence contour plot for MJD = 57188.77, for (a) $\dot{m}_d$-$\dot{m}_h$ and (b) $X_s$-$R$.
}
\end{figure}

\begin{figure}
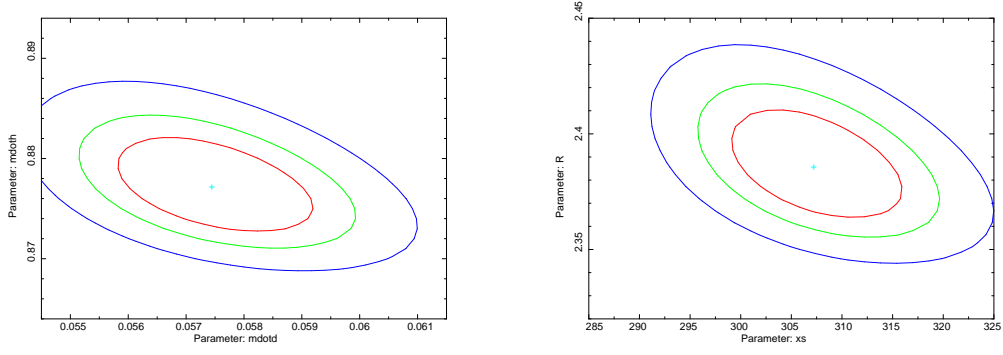

 \hbox{
       \hspace*{0.2cm}\includegraphics[angle=270,width=7cm]{md-mh-cntr-xrt.ps}
       \hspace*{0.2cm}\includegraphics[angle=270,width=7cm]{xs-r-cntr-xrt.ps}
        }
 \caption{2D confidence contour plot for MJD = 57189.62, for (a) $\dot{m}_d$-$\dot{m}_h$ and (b) $X_s$-$R$.
}
\end{figure}

\section{Results}
\label{result}

We present the results of spectral and temporal analysis of the source in $0.5-150$~keV energy band using 
combined XRT+BAT or only BAT (in $15-150$~keV) or only XRT (in $0.5-10$~keV) data. In Fig. 4, we show two 
XRT PDS for observation on MJD = 57191.01 (2015 June 18) and MJD = 57194.54 (2015 June 21). In Fig. 5, 
TCAF model fitted spectrum of combined XRT plus BAT data in the broad energy range $0.5-150$~keV is shown
for the observation on MJD = 57191.01. 

In Fig. 6a, we show the evolution of BAT and XRT fluxes. In Fig. 6(b-c), we show the variation of the
Keplerian disk rate ($\dot{m_d}$), the sub-Keplerian halo rate ($\dot{m_h}$) with day (in MJD). In Fig. 
4d, we show the evolution of accretion rate ratio (ARR = $\dot{m_h}$/$\dot{m_d}$). In Fig. 7a, we show 
the variation of the equipartition magnetic field with the day (see, below for details). In Fig. 5(b-d), 
we show the variation of the shock location ($X_s$), the shock compression ratio ($R$) and TCAF model 
normalization with the day. In Fig. 8, the time-resolved {\it Swift}/BAT spectra in the energy range of 
$15-150$ keV for the observation on June 15, 2015 (MJD = 57188.77) are shown. The three spectra are
marked as (a) (online green), (b) (online black) and (c) (online red) with exposures time of 220 sec,
680 sec and 160 sec, respectively.

\begin{figure}
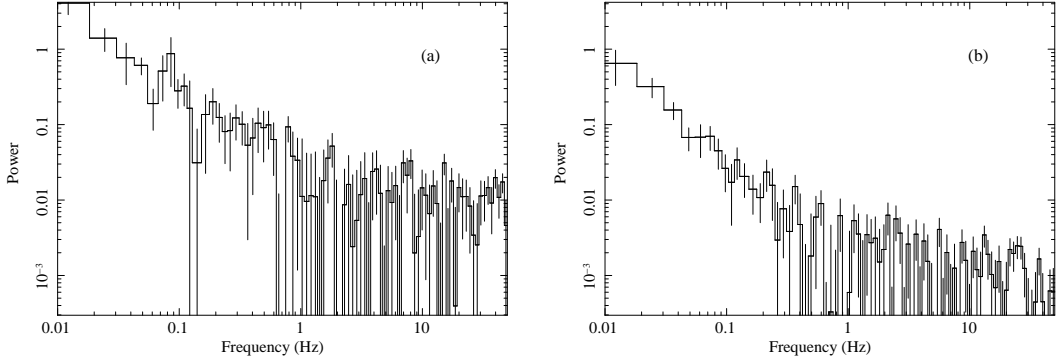

 \hbox{
       \hspace*{0.2cm}\includegraphics[angle=270,width=7cm]{fig2a.ps}
       \hspace*{0.2cm}\includegraphics[angle=270,width=7cm]{fig2b.ps}
        }
 \caption{Power density spectra (PDS) of $0.01$ sec time binned lightcurves of $0.5-10$~keV XRT data. 
The observation date of these PDS are: 2015 June 18 (MJD = 57191.01) and 2015 June 21 (MJD = 57194.54).
}
\end{figure}

\subsection{Temporal Evolution}
\label{timing}

The 2015 outburst of V404 Cygni is not like any other regular outburst of transient BHCs. It showed 
rapid changes in very short timescale in both XRT and BAT count rates. The flux changed significantly 
in minute to hour timescales. Several radio and X-ray flares were observed during this epoch. {\it INTEGRAL}
observation revealed $18$ X-ray flares during the outburst \citep{Rodriguez15}. Radio flare were also 
reported \citep{Mooley15a,Trushkin15b,Tetarenko17}. We find that the luminosities were Eddington or 
super-Eddington in some observations. On 2015 June 22 (MJD = 57195.41), the XRT flux increased rapidly 
to $4.61\times 10^{-7}$ ergs s$^{-1}$ from $8.2\times 10^{-8}$ ergs s$^{-1}$ of the previous observation. 
This corresponds to the source luminosity, $L \simeq 3.13 \times 10^{38}$ ergs sec$^{-1}$. On June 25, 2015 
(MJD = 57198.93), the XRT flux reached at $4.59\times 10^{-6}$ ergs sec$^{-1}$ which corresponds to the source 
luminosity, $L \simeq 3.12 \times 10^{39}$ ergs sec$^{-1}$. BAT count rate also increased rapidly within our
analysis period. It became maximum on June 26, 2015 (MJD = 57198.93) at $4.68 \times 10^{-7}$ ergs sec$^{-1}$, 
$i.e.,$ luminosity was $L \simeq 3.18 \times 10^{38}$ ergs sec$^{-1}$. 

We studied power density spectra (PDS) of $0.01$~sec binned lightcurve of {\it Swift}/XRT. We did not 
find any clear evidence of QPOs during the initial phase of the outburst in both XRT and BAT PDS. A weak 
signature of QPO was observed in the PDS on 2015 June 18 (MJD = 57191.03) (see. Fig 4a). The XRT PDS
showed almost flat spectral slope with broad-band noise. The noise decreased as the outburst progressed
during which we found two slopes in the PDS, steep powerlaw slope in the lower frequency and flat slope 
at a higher frequency. In some PDS, we observed that power diminished very rapidly. We did not find any
break in the PDS. Similar nature is also observed in BAT PDS.

\begin{figure}
\begin{center}
\includegraphics[angle=270,width=8.5cm,keepaspectratio=true]{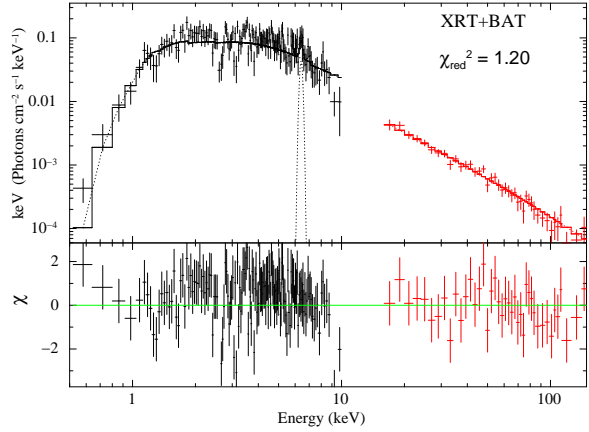}
\end{center}
\caption{TCAF model fitted combined XRT+BAT spectra in the energy range of $0.5-150$ keV, observed on MJD = 57191.03.}
\label{5}
\end{figure}

\subsection{Evolution of the Spectral Properties}
\label{specral}

We have done the spectral analysis using $0.5-150$~keV combined {\it Swift}/XRT and {\it Swift}/BAT 
data between 2015 June 15 and 26. \citep{Radhika16} analyzed the same data set using phenomenological
{\it diskbb} and {\it powerlaw} models. We analyzed the data with combined {\it diskbb} and {\it powerlaw}
models and have found similar results as in \citep{Radhika16}. In general, we used `$phabs*pcf*(diskbb+powerlaw+gaussian)$'
model to estimate thermal and non-thermal fluxes. While analyzing with phenomenological models, we did
not require diskbb component on a regular basis. {\it Diskbb} component was required only in 9 observations 
out of a total 19 observations. During the entire period, PL photon index varied between $0.60$
and $2.43$ although the PL flux dominated over the DBB flux. We also required a `$Gaussian$' for iron 
line emission along with the `$diskbb+powerlaw$' model. Detailed results of the phenomenological model are given 
in Table I.

In the present paper, our main goal is to study the accretion flow dynamics of the source from spectral
analysis with the physical TCAF model. For this purpose, we used `$phabs*pcf*(TCAF+gaussian)$' model. 
Detailed results using this model is presented in Table II. From the spectral analysis in the initial 
few observations, we obtained very low values of the Keplerian disk rate ($\dot{m_d}$) while the 
sub-Keplerian halo rates ($\dot{m_h}$) were found to be high ($> 0.85 \dot{M}_{Edd}$). On 2015 June 19 
(MJD = 57192.16), $\dot{m_h}$ increased to $1.37 \dot{M}_{Edd}$ from its previous day value of 
$0.85 \dot{M}_{Edd}$. After that, it varied within $1.24-1.58$ $\dot{M}_{Edd}$ until the end of our 
analysis period. On 2015 June 25 (MJD = 57198.02), we observed sudden rise in $\dot{m_d}$ from its 
previous day, i.e., from $0.11 \dot{M}_{Edd}$ to $0.15 \dot{M}_{Edd}$. Before that, $\dot{m_d}$ varied 
in the range of $0.05-0.12$ $\dot{M}_{Edd}$. After MJD = 57198.02, $\dot{m_d}$ was obtained in a narrow 
range of $0.13-0.16$ $\dot{M}_{Edd}$ (see Fig. 6b).
 
A strong shock ($R=2.56$) was found far away from the black hole ($X_s=334~r_s$) on the first day
(MJD=57188.77) of our observation (see Fig. 7b \& 7c). The shock remained strong for the next 
five days. After that, the shock was found to move closer to the black hole as the Keplerian disk 
rate increased. The shock was found at $126$ $r_s$ on MJD = 57194.16 with $R=1.92$. The shock did 
not move closer than this. After that, shock moved away from the black hole. Again we found that 
the shock was moving inward after MJD=57197.21. On the last day of our observation, we found the 
shock to be at $143$ $r_S$. 

We add a $Gaussian$ profile along with the TCAF solution to incorporate the contribution of the $Fe$ 
emission line in XRT data. Fe-line varied within $6.08$~keV and $6.97$~keV. In some observation, 
we required the line width of $>1$~keV. We used $phabs$ models for the interstellar absorption. 
We did not freeze $n_H$ at a particular value. Rather, we kept it free. In our analysis, we observed
it to vary between $0.54 \times 10^{22}$ to $1.49 \times 10^{22}$ cm$^{-2}$. We also used $pcfabs$
model to incorporate for partial absorption in XRT data. In some observations, the covering required
as high as 95\%. In general, it varied between 50\% and 95\%. For covering absorption, $n_H$ varied 
between  $2.2 \times 10^{22}$ and $28.9 \times 10^{22}$ cm$^{-2}$.

\begin{figure}
\begin{center}
\includegraphics[width=8cm,keepaspectratio=true]{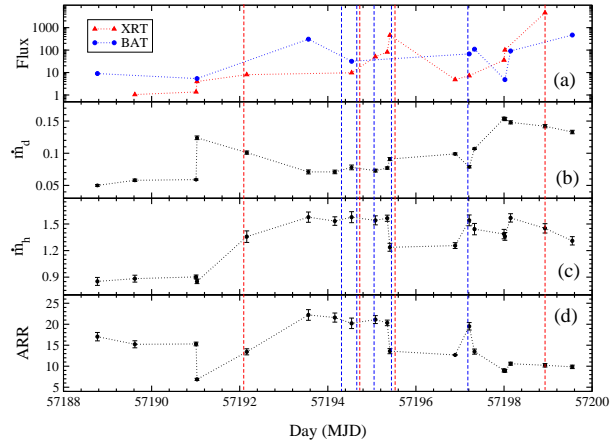}
\end{center}
\caption{Panel (a) shows variation of $0.5-10$~keV XRT and $15-150$~keV BAT flux with day (MJD). 
The red (online) triangle represents XRT flux while blue (online) circle represents BAT flux. The fluxes are 
in the unit of $10^{-9}$ $ergs/sec$. The variation of (b) $\dot{m_d}$ in $\dot{M}_{Edd}$ and (c) $\dot{m_h}$
in $\dot{M}_{Edd}$ are shown with day (MJD). In panel (d) the variation of ARR (=$\dot{m_h}/\dot{m_d}$) is shown.
Blue and red dotted lines represent reported X-ray and Radio flaring activities respectively.}
\label{5}
\end{figure}

\begin{figure}
\begin{center}
\includegraphics[width=8cm,keepaspectratio=true]{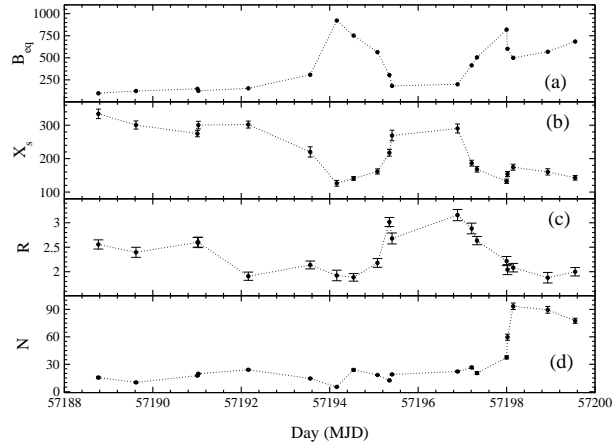}
\end{center}
\caption{The variation (a) equipartition magnetic field ($B_{eq}$) in $Gauss$, (b) shock location ($X_s$) in $r_s$, 
(c) shock compression ratio ($R$) and (d) normalization ($N$) are shown with day (MJD).}
\label{6}
\end{figure}

\subsection{Time Resolved BAT Spectra}
\label{bat}
To study the evolution of spectral nature in short time intervals, we analyzed the time-resolved BAT spectra.
Rapid variation of the accretion rates and other physical flow parameters from observation to observation 
motivated us to make this study. Here, we analyzed time-resolved BAT spectra for four observations on June 15, 
2015 (MJD = 57188.77), June 18, 2015 (MJD = 57191.03), June 20, 2015 (MJD = 57193.56) and June 26, 2015 
(MJD = 57199.52). We found rapid variation in the BAT spectra within very short period of time even in one 
observation. In Table III, TCAF model fitted parameters for time-resolved spectra are presented. For example,
on 2015 June 15, within total BAT exposure of 1202 sec (Fig. 8), we observed the variations of $\dot{m_d}$ 
between $0.041$ and $0.057$ $\dot{M}_{Edd}$, and $\dot{m_h}$ between $0.82$ and $0.88$ $\dot{M}_{Edd}$.
The shock was observed to vary in between $\sim 326-336~r_s$. Similar rapid variation of the flow parameters
($\dot{m_d}$ \& $\dot{m_h}$) were also observed for the remaining three observations.

\begin{figure}
\begin{center}\includegraphics[angle=270,width=8.5cm,keepaspectratio=true]{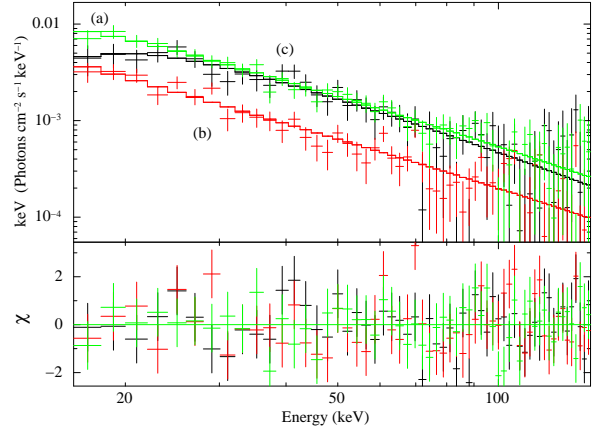}
 \end{center}
\caption{Time resolved BAT spectra for observation on MJD = 57188.77 is shown. The spectra correspond
to exposures of (a) 220 sec, (b) 680 sec and (c) 160 sec.}
\label{7}
\end{figure}

\subsection{Estimation of Magnetic Field}
Observation of presence of high magnetic field on 2015 June 25 by \citep{Dallilar17} motivated us to estimate 
magnetic field strength at the `hot' Compton cloud region (here CENBOL) for the BHC V404 Cygni during its 2015 
outburst. We made some simple assumptions as mentioned below to calculate the equipartition value of the magnetic
field. At the shock location, energy conservation leads to the following equation \citep{C90},

\begin{equation}
na_{+}^2 + \frac{1}{2}V_{+}^2 = na_{-}^2 + \frac{1}{2}V_{-}^2 ,
\end{equation}
where, $n$ is the polytropic index of the flow, $a$ is sound speed, $V$ is particle velocity. $`+'$ and $`-'$ 
signs indicate the values at post- and pre-shock region respectively. The electron number density ($n_e$) is 
given by, 
\begin{equation}
    n_e = \frac{\dot{m_d}+\dot{m_h}}{4 \pi X_s V_{+} H_{shk} m_p} .
\end{equation}

For our calculation, we assume CENBOL shape as cylindrical. $m_p$ is the mass of the proton. $H_{shk}$ is 
the shock height. The shock height could be calculated from the TCAF model fitted shock parameters \citep{DMC15}
using standard vertical equilibrium \citep{C89},

\begin{equation}
H_{shk}=\sqrt{\frac{n(R-1)X_s^2}{R^2}} .
\end{equation}
Now, we can calculate pressure at CENBOL using the following equation,
\begin{equation}
P_{gas}=\frac{a_{+}^2 m_e n_e}{n} .
\end{equation}

We consider equipartition magnetic field ($B$) as, 
\begin{equation}
\frac{B^2}{8\pi} = P_{rad}+P_{gas} ,
\end{equation}

where, $P_{rad}$ is the radiation pressure. The radiation pressure is negligible compared to the gas pressure. 
Using this equation, we calculated the maximum possible value of the equipartition magnetic field for this source 
(see, Table IV). We show the variation of the equipartition magnetic field in Fig. 7a. We find $B = 97~G$ on 
our first observation (MJD = 57188.77). Then we found that $B$ was increasing with the shock moving towards 
the black hole. $B$ was maximum on 2015 June 21 (MJD = 57194.16) at $923~G$. On that day, the shock was closest
to the black hole. 

\subsection{Cooling Time}
\label{cooling}

V404 Cygni did not show any strong QPO during its 2015 outburst. Although there are numerous suggestions 
for the origin of low frequency QPOs, we believe that these QPOs are generated due to the oscillation of
the shock formed in TCAF solution. Ths shock oscillates when the resonance condition between cooling and 
infall time scales is satisfied or when Rankine-Hugoniot conditions are not satisfied to form a stable 
shock \citep{MSC96,Ryu97,C15}. Thus to check if the resonance condition is satisfied or not, we have 
calculated both cooling and infall time scales during the outburst. Generally in low mass X-ray binaries, 
Compton cooling is the primary process of cooling. We also considered the synchrotron cooling since the 
magnetic field was present.

We calculated both cooling time scales (synchrotron and Compton) during the entire period of our observations 
and compared that with the infall times. If cooling and infall timescales are roughly comparable, then we may 
say that the resonance condition for oscillation of the shock is satisfied. For simplicity, here we assume that 
the matter is moving radially at the post-shock region with speed $V_{+}$. This allowed us to calculate 
infall time using the following equation,

\begin{equation}
t_{inf} = \frac{X_s}{v_{+}}
\end{equation}

Similar to the magnetic field calculation, here we also assume CENBOL as a cylindrical in shape. Now the total
thermal energy of electron content within the CENBOL is,

\begin{equation}
E=\gamma_e m_e c^2 V n_e ,
\end{equation}
where $\gamma_e$ is the Lorentz factor. It is given by, $\gamma_e = 1/\sqrt{1-\beta_e^2}$,  where $\beta_e=v/c$, 
$v$ being electron velocity and $c$ is the velocity of light. The synchrotron cooling rate is \citep{Rybicki79},

\begin{equation}
\Lambda_{syn}=\frac{4}{3}\sigma_T c \beta_e \gamma_e^2 U_B ,
\end{equation}
where $\sigma_T$ is Boltzman constant. $U_B$ is magnetic energy density and given by, $U_B=B^2/8 \pi$. 
The synchrotron cooling time is given by, 

\begin{equation}
t_{syn}=\frac{E}{\Lambda_{syn}}.
\end{equation}

We calculated cooling time due to synchrotron using Eqn. 9. Note, we use $\gamma_e$=10 for our calculation.
We also checked the Compton cooling timescale ($t_{Comp}$). To calculate Compton cooling, we use the same
method as described in C15. The Compton cooling timescale is given by Eqn. 6 of C15,

\begin{equation}
 t_{Comp} = \frac{E}{\Lambda_{Comp}}.
\end{equation}

Here, $\Lambda_{Comp}$ is Compton cooling rate (for more details, see, C15). Here we find that the Compton 
cooling is much faster than the synchrotron cooling (see, Table IV). Thus we may assume that inverse-Comptonization 
is the primary process for cooling. Now if we compare $t_{inf}$ with $t_{syn}$ or with $t_{Comp}$, in both 
the cases, the ratio deviates largely from unity. Thus we may say that during the 2015 outburst of V404~Cygni, 
resonance conditions are not satisfied to form oscillating shock. This indeed verifies the observational fact
of the no prominent signature of type-C low frequency QPOs during the outburst. The detailed results of our 
calculation of the magnetic field and two types of cooling time scales are presented in Table IV.

\section{Discussion}
\label{discussion}

{\it Swift}/XRT and BAT data for 19 observations between 2015 June 15 (MJD=57188.77) and June 26 (MJD=57199.52) 
are used to study properties of the V404 Cygni in a broad energy range of $0.5-150$~keV. The TCAF model fitted 
spectral analysis allowed us to understand the nature of this violent outburst from a physical perspective. The 
source exhibited this outburst after a long $26$ years of quiescent. During this outburst, many radio and X-ray 
flares were also observed. The presence of a strong magnetic field is considered to be one of the reasons behind
this unusual outburst of V404~Cygni. In the following sub-Sections, we discuss those in details.

\subsection{Magnetic Field}

A magnetic field is brought in by the accretion flow. The shear, convection and advection amplify the 
predominantly toroidal flux tubes. They are expelled from the CENBOL due to the magnetic buoyancy 
\citep{CD94,DC94} in the direction towards the pressure gradient force. If the buoyancy timescale is 
larger than the shear amplification timescale, then the magnetic flux tubes are amplified within the 
dynamical timescale until they reach equipartition where gas pressure matches with the magnetic pressure. 
In general, the magnetic field is not found to contribute to the accretion disk spectra of black holes
since TCAF alone fits the data very well. However, it is necessary for acceleration and collimation of
jets and outflow.

Here, we calculated the equipartition magnetic field by equating magnetic pressure with the local gas 
pressure at the shock. We found $B_{eq}=97$~$G$ on the first observation day. After that, it increased 
gradually with the increase of the accretion rate. We found the maximum value of the magnetic field to
be $B_{eq}=923$ $G$ on June 21, 2015 (MJD = 57194.16). Then, it decreased to $154$ G on MJD = 57196.89. 
Then, we observed that it was varying between $414$ and $820$ $G$ within our observation. Dallilar et al.
calculated the equipartition magnetic field for this source to be $461\pm 12$ $G$ on June 25, 2015
(MJD = 57198.18) \citep{Dallilar17}. We found it $B_{eq}=500 \pm 18$ $G$ on that day (MJD = 57198.15). 
However, our estimated value of the magnetic field is much lower than the previously estimated values 
for other Galactic black holes. Previously, magnetic field was found to be about $\sim 10^5 - 10^7$ $G$ 
for Cygnus X-1 \citep{Del Santo13}; $\sim 5 \times 10^4$ $G$ for XTE J1550-564 \citep{Chatty11}; and 
$\sim 1.5 \times 10^4$ $G$ for GX~339-4 \citep{Cutri03}. However, the magnetic field was calculated in a 
jet for these sources; thus, they showed much higher values.

\subsection{Power Density Spectra}

No prominent QPOs were observed in the PDS of the Fourier transformed $0.01$~sec time binned XRT lightcurves
in the $0.5-10$~keV energy band. Presence of a significant magnetic field may cause turbulence in the accretion 
disk, which could be responsible for the white noise observed in the PDS. The PDSs were observed to have two 
slopes: a powerlaw component and a flat component. Powerlaw slope was observed in the lower frequency region 
while flat slope was observed in the higher frequency. The turbulence could be the reason behind the flat
spectrum observed in the XRT PDS. Power in the PDS follows as, $P \sim 1/f^{\beta}$, where $f$ is the frequency 
and $\beta$ is powerlaw index. In the most PDS, we found $\beta \sim 0$, $i.e.$, flat spectra. It indicates 
the presence of turbulence in the Keplerian disk. No break was found in power density spectra. This also 
indicates no or minimal contribution of the Keplerian disk accretion. Precisely, this is found from the 
spectral analysis.

\subsection{Absence of QPOs}

We believe that the oscillation of shock is responsible for the QPOs \citep{MSC96}. Generally strong type-C 
QPOs are observed if the resonance condition is satisfied, i.e., when the infall time of the post-shock matter
roughly matches with the cooling time inside the CENBOL. The shock oscillation could also be observed when 
Rankine-Hugoniot conditions are not satisfied to form a stable shock \citep{Ryu97}. According to C15, the 
type-B or type-A QPOs mainly occur due to weak resonance phenomenon in CENBOL (type-B) or in shock-less
centrifugal barrier (type-A). The cooling process could be via inverse-Compton scattering, bremsstrahlung or 
synchrotron emission. If the magnetic field is high enough, one can expect the dominance of synchrotron cooling.

We did not observe any prominent type-C QPOs during the 2015 outburst of V404 Cygni. Non-satisfaction of the
resonance condition could be responsible for it. To verify this assertion, we calculated synchrotron and Compton
cooling times for this source. We found that the Compton cooling rate is much faster than the synchrotron cooling
rate. Thus Compton cooling dominated the cooling process. We compared both types of cooling times with the infall
time. We found that the infall time and cooling time were not comparable at all (see Table IV). C15 showed that
QPO would be generated if the ratio of the infall and the cooling times is between $0.5$ and $1.5$, $, i.e.,$
within 50\% of unity either way. Since the resonance condition was not satisfied, we were not supposed to see 
any strong type-C QPOs.

Huppenkothen et al. 2017 reported of detection of mHz QPO with {\it Chandra}, {\it Swift}/XRT and {\it Fermi}
observations \citep{Huppenkothen17}. They reported simultaneous detection of $18$~mHz QPO with {\it Swift}/XRT
and {\it Fermi}/GBM. They classified this QPO as a new type of low-frequency QPO. However, Radhika et al. argued 
that they did not find any clear signature of this QPO \citep{Radhika16}.  {\it Chandra}/ACIS observation revealed 
signatures of $73$~mHz and $1.03$~Hz QPOs. However, they did not seem to be type-C QPOs. Thus the resonance 
condition was not behind the origin of these QPOs. This could be due to non-satisfaction of the Rankine-Hugoniot 
conditions at the shock front.

\subsection{Spectral and Temporal Evolution}

V404 Cygni showed complex behaviour during the 2015 outburst. We required two absorption models to fit the spectra: 
$phabs$ model for interstellar absorption and $pcfabs$ for the partial covering absorption. The later absorption
may be due to disk wind emission or outflow emission \citep{King15,Radhika16}. Variable dust scattering rings were 
observed with XRT \citep{Beardmore et al. 2015,Vasilopoulos16}. After MJD = 57198, the dust halo started to dominate 
in the field of view; hence we studied the spectral and the timing properties up to this day. 

Some observations of V404 Cygni could not be fitted with simple disk blackbody and powerlaw models since a significant 
reflection component was present. In the first six observations, DBB component was not required to fit the spectra 
(till MJD = 57193.56). After that, it was required occasionally. PL photon indices varied randomly between $0.60$ and 
$2.43$. Overall, PL flux dominated over the thermal flux during the entire period of our observation. We often observed 
a `hump' region after $\sim 8$ keV in the XRT data. This `reflection hump' was extended up to $\sim 20$ keV in the BAT
data. Radhika et al. had also analyzed $0.5-150$~keV combined XRT+BAT data using disk blackbody (diskbb) and PL models
and found similar results \citep{Radhika16}. Later, they used `pexrav' model instead of `PL' model for this reflection.
In their analysis, disk blackbody component was not required in a few observations. They found photon indices varied
randomly between $0.60$ and $4.43$. \citep{Rodriguez15,Natalucci15} and \citep{Roques15} analyzed {\it INTEGRAL}/IBIS
and {\it INTEGRAL}/SPI data with Comptonization model in the broad energy range of $20-650$ keV. They required an 
additional cut-off powerlaw in the energy range of $400-600$~keV. They found that Comptonization temperature was 
$\sim 40$~keV with seed photon temperature $\sim 7$~keV. This is very high for the disk emission. They concluded that 
this emission could be of different origins, such as synchrotron emission from the jet.

We extracted the physical parameters of the accretion flows from each fit. We first fitted each spectrum by keeping 
all model input parameters, including the mass of the black hole as free. We found a variation of $M_{BH}$ in a narrow
range of $9.5-11.5~M_{\odot}$ with an average value of $10.6^{+0.9}_{-1.1}$ $M_{\odot}$. This estimated mass of V404
Cygni agrees well with other reported values in the range of $9-12$ $M_{\odot}$ \citep{Casaries1992,Shahbaz94,Khargharia10}. 
We then re-fitted all the spectra by keeping $M_{BH}$ frozen at its most probable value ($=10.6$~$M_{\odot}$) to extract
values of other physical flow parameters during the outburst.

From the variation of the TCAF model fitted flow parameters (i.e., high dominance of $\dot{m_h}$ over $\dot{m_d}$
and presence of strong shock ($R>2.3$) at a large distance), we infer that the source was in the hard state at the
beginning of the outburst. High ARR ($\sim 14-22$) also indicates this except in one observation made on 2015 June
18 (MJD=57191.03). From the phenomenological model fittings, the dominance of powerlaw flux also indicates this. On 
2015 June 25 (MJD=57198), we observed that $\dot{m_d}$ started to increase rapidly although other parameters did not 
change much, leading to softening of the spectra. At this last phase of our observation, we observed that ARR was 
decreasing with an increase in the Keplerian disk rate. During this phase, the source may be in the hard-intermediate
state. In general, spectral state classification is done based on the variation of ARR and the nature of QPO. Due to
the absence of QPOs and complex behaviour of the ARR, we have not been able to assert the exact state transition day 
but the source likely entered in the hard-intermediate state on MJD = 57197.21 when $\dot{m_h}$ was found to achieve 
its maximum rate. After this day, $\dot{m_d}$ was found to increase rapidly. However, our classification may not be 
valid since the source showed rapid fluctuation in very short timescale \citep{Motta17}. Motta et al. analyzed the 
time-resolved {\it Swift}/XRT spectra and found that spectra shape changed in as short as $\sim 10$ sec \citep{Motta17}.

The 2015 outburst is somewhat similar to the 2003 outburst of H~1743-322. Both outbursts occurred after long quiescence 
periods. Like the 2015 outburst of V404 Cygni, the 2003 outburst of H~1743-322, which also occurred after about 26 years,
showed peculiar behaviour with several flares and outflow activity. Thus it is possible that a huge amount of matter
accumulated at a large pileup distance over a long time. With a sudden enhancement of viscosity, the outburst is triggered
and leads to a violent and non-settling activity \citep{Chakrabarti19}.

\subsection{Evolution of the Spectral Properties with Flares}

The present outburst of V404 Cygni did not behave like any other typical outburst of a classical transient black hole. 
V404 Cygni showed a strong jet associated with several flares. The flares were observed in multi-wavebands, from X-ray, 
optical, IR to radio \citep{Rodriguez15,Gandhi16,Trushkin15a,Trushkin15b,Tetarenko17}. Eighteen X-ray flares were 
reported with the {\it INTEGRAL} and {\it SWIFT}/BAT observations between June 20, 2015 (MJD = 57193) and June 25, 2015 
(MJD = 57198). The magnetic field was the strongest during this phase of the outburst. The magnetic field could be 
responsible for this flaring activity.

In general, we see a decreasing ARR in an observation immediately after a flare, indicating softening of the spectra
(see Fig. 6). For example, an X-ray flare was observed on MJD = 57194.31 \citep{Rodriguez15}. This could be due to 
the high magnetic field on MJD = 57194.16. Immediately after the flare, on MJD = 57194.54, we found that the ARR 
decreased slightly from its previous observation ($21.6$ to $20.2$). On June 18, 2015 (MJD = 57191) we observed that
the Keplerian disk rate suddenly rose to $0.12$~$\dot{M}_{Edd}$ from $0.06$ $\dot{M}_{Edd}$ within $\sim 30$ mins.
This could be associated with the radio flare observed on MJD = 57191.09 with AMI-LA observation \citep{Mooley15a}. 
This is expected since a large amount of mass was ejected from the CENBOL during a flare and inflowing matter rapidly 
moved inward to fill the vacant space. This led to the softening of the spectrum as the CENBOL size was reduced. 
However, this was not observed after every flare. It is possible that those flares were not localized and the disk
was unstable. Around MJD = 57198, we found that $B_{eq}$ decreased sharply, although $N$ increased very rapidly. This 
could be due to the high magnetic field which produced flare and outflow. This flare and the outflow was responsible
for the rapid rise of $N$.

The X-ray jet flux can be calculated based on the deviation of the constancy of the TCAF model normalization 
\citep{JCD17,Jana20a,C19}. However, to calculate jet X-ray flux by this model, we must have at least one observation
where the effects of the jet were negligible. In that observation, the entire observed X-ray should be contributed 
only from the inflowing matter of the accretion disk and CENBOL. However, for V404 Cygni, a strong jet was present in
all the observations. Thus we were not able to separate the jet X-ray contribution from the total X-ray. Random variation
in normalization may be due to the presence of fluctuating magnetic field or unsettling disk, which led to the flaring
activity of the source.

\section{Summary}
\label{summary}

The first epoch of the 2015 outburst of the Galactic black hole V404 Cygni was an unusual and violent outburst.
It did not behave like other typical outbursts of Galactic transient black hole candidates. Rapid variations 
were observed in both spectral and timing properties in a very short time scales, ranging from a few minutes 
to hours. We have used $0.5-150$~keV combined {\it Swift}/XRT and {\it Swuft}/BAT data to study the accretion 
flow properties of the source. Spectral analysis was done using the TCAF model-based {\it fits} file in {\tt XSPEC}. 
The model fitted/derived flow parameters allowed us to understand the evolution of accretion flow parameters of 
this violent outburst. We have also calculated the equipartition magnetic field for the outburst. No break is 
found in the power density spectra, which indicates that the Keplerian disk rate was very low. This is also 
confirmed form the spectral analysis. The presence of white noise in higher frequencies in the power density
spectra indicates the presence of a highly turbulent disk. The strong magnetic field could be the reason behind it. 
It is also responsible for the flares. We find that the Compton cooling process is much faster than the synchrotron
cooling process. Since the resonance condition between cooling and infall time scales inside the CENBOL is not satisfied, 
we did not expect any sharp low frequency QPO. Indeed the object did not show any signature of prominent type-C QPO.

\begin{acknowledgements}

This work made use of XRT and BAT data supplied by the UK {\it Swift} Science Data Centre at the University of Leicester. 
We acknowledge anonymous referee for his kind suggestion to improve the quality of the paper.  
A.J. and D.D. acknowledge support from DST/GITA sponsored India-Taiwan collaborative project (GITA/DST/TWN/P-76/2017) fund.
A.J. also acknowledges CSIR SRF fellowship (09/904(0012) 2K18 EMR-1).
Research of D.D. and S.K.C. is supported in part by the Higher Education Dept. of the Govt. of West Bengal, India.
D.D. also acknowledges the ISRO sponsored RESPOND project (ISRO/RES/2/418/17-18) fund.
D.C. and D.D. acknowledge support from DST/SERB sponsored Extra Mural Research project (EMR/2016/003918) fund.
J.-R. S., and H.-K. C. are supported by MOST of Taiwan under grants MOST/106-2923-M-007-002-MY3 and MOST/107-2119-M-007-012.
\end{acknowledgements}

\clearpage
% \begin{landscape}
\begin{table}
\vskip 0.5cm
\addtolength{\tabcolsep}{-4.50pt}
\scriptsize
\centering
\centering{\large \bf Table I: DBB+PL Model Fitted Spectral Analysis Results}
\vskip 0.2cm
\begin{tabular}{lcccccccccccccc}
\hline
Obs ID&Day&XRT exp&BAT exp&XRT Flux$^*$ & BAT Flux$^*$&DBB Flux$^*$&PL Flux$^*$&${n_H}^1$&${n_{H}}^2$& $CF$ &$T_{in}$ &$\Gamma$&$\chi^2 /dof$\\
  &($MJD$)&(sec) &(sec)& &  &  &&$(\times 10^{22})$  & $(\times 10^{22})$ & & $(keV)$& & &\\
 (1)&  (2)  & (3)  & (4)& (5) & (6) & (7) & (8) & (9)& (10)& (11)&(12) &(13)& (14)  \\
\hline

00643949000&  57188.77&--- &1202&   ---             &$  9.01 ^{\pm0.09}$ &        ---         &$  9.01^{\pm0.09}$  &$0.94^{\pm0.07}$&    ---         &    ---         &  ---           &$1.59^{\pm0.13}$&   71/53  \\ 
00031403035&57189.62&1970&--- &$  1.04^{\pm0.04}$ &   ---              &        ---         &$  1.04^{\pm0.04}$  &$1.12^{\pm0.09}$&$32.7^{\pm0.2}$ &$0.46^{\pm0.06}$&  ---           &$1.09^{\pm0.15}$& 714/944 \\
00031403038 & 57191.01&610 &--- &$  1.36^{\pm0.06}$ &   ---               &      ---         &$  1.36^{\pm0.06}$  &$0.91^{\pm0.09}$&$1.61^{\pm0.06}$&$0.82^{\pm0.10}$&  ---           &$0.60^{\pm0.07}$& 1013/946\\
00644520000 & 57191.03&217 &1202&$  3.98^{\pm0.07}$ &$  5.37 ^{\pm0.08}$ &        ---         &$  9.35^{\pm0.15}$  &$0.59^{\pm0.08}$&$10.5^{\pm0.4}$ &$0.73^{\pm0.12}$&  ---           &$1.83^{\pm0.18}$&  394/250\\
00031403042 & 57192.16&1262&--- &$  8.05^{\pm0.03}$ &   ---              &        ---         &$  8.05^{\pm0.03}$  &$1.28^{\pm0.05}$&$11.7^{\pm0.6}$ &$0.95^{\pm0.10}$&  ---           &$1.39^{\pm0.18}$& 1339/931\\
00645176000 & 57193.56&--- &542 &    ---            &$  306.7 ^{\pm2.3}$ &        ---         &$ 306.7^{\pm2.3} $  &$0.51^{\pm0.03}$&    ---         &    ---         &  ---           &$1.20^{\pm0.15}$&  56/51  \\
00031403048 & 57194.16&4208&--- &$  4.89^{\pm0.08}$ &   ---              &$  1.42^{\pm0.03}  $&$  3.47^{\pm0.04}$  &$0.49^{\pm0.03}$&    ---         &    ---         &$0.19^{\pm0.03}$&$1.65^{\pm0.17}$& 970/889 \\
00031403046 & 57194.54&240 &7   &$  9.77^{\pm0.10}$ &$  31.8 ^{\pm0.7}$  &        ---         &$   41.6^{\pm0.3}$  &$0.66^{\pm0.07}$&    ---         &    ---         &  ---           &$1.07^{\pm0.19}$& 1437/882\\
00031403045 & 57195.08&930 &--- &$  50.8^{\pm1.0}$  &   ---              &$   24.0^{\pm0.4}  $&$   26.7^{\pm0.6}$  &$0.54^{\pm0.05}$&$5.97^{\pm0.14}$&$0.88^{\pm0.12}$&$0.15^{\pm0.03}$&$0.44^{\pm0.06}$& 1192/928\\
00031403049 & 57195.35&2977&--- &$  82.1^{\pm1.5}$  &   ---              &$   16.9^{\pm1.5}  $&$ 65.2^{\pm2.2}  $  &$0.62^{\pm0.04}$&$0.44^{\pm0.02}$&$0.69^{\pm0.07}$&$0.95^{\pm0.12}$&$0.55^{\pm0.06}$& 1282/702 \\
00031403047 & 57195.41&1903&--- &$  461.9^{\pm3.4}$ &   ---               &        ---         &$  461.9^{\pm3.4}$  &$0.71^{\pm0.05}$&    ---         &    ---         &  ---           &$1.25^{\pm0.14}$&  975/713\\
00031403052 & 57196.89&275 &--- &$  4.84^{\pm0.09}$ &   ---               &$   2.67^{\pm0.21} $&$  2.16^{\pm0.16}$  &$0.75^{\pm0.08}$&$5.83^{\pm0.15}$&$0.95^{\pm0.05}$&$0.24^{\pm0.07}$&$1.48^{\pm0.29}$& 1061/941\\
00031403054 & 57197.21&682 &5   &$  7.21^{\pm0.15}$ &$  67.8 ^{\pm1.2}$  &$   1.04^{\pm0.08} $&$  74.0^{\pm1.1} $  &$0.96^{\pm0.11}$&$29.4^{\pm1.2}$ &$0.77^{\pm0.11}$&$0.21^{\pm0.06}$&$1.43^{\pm0.22}$& 1409/979\\
00031403053 & 57197.33&1052&518 &$  7.33^{\pm0.12}$ &$  109.7 ^{\pm1.5}$ &        ---         &$ 117.0^{\pm1.5} $  &$0.48^{\pm0.09}$&    ---         &    ---         &  ---           &$2.43^{\pm0.36}$& 1107/998\\
00031403055 & 57198.00&1028&--- &$  35.9^{\pm0.4}$  &   ---              &$   19.7^{\pm0.8}  $&$ 16.2^{\pm0.8}  $  &$1.08^{\pm0.18}$&$0.91^{\pm0.11}$&$0.61^{\pm0.09}$&$0.76^{\pm0.10}$&$1.67^{\pm0.24}$& 1250/926\\
00031403056 & 57198.02&818 &713 &$  101.6^{\pm1.8}$ &$  4.82 ^{\pm0.22}$ &$  101.6^{\pm1.8}  $&$  4.82^{\pm0.33}$  &$0.58^{\pm0.08}$&$9.15^{\pm0.81}$&$0.81^{\pm0.11}$&$0.17^{\pm0.03}$&$1.76^{\pm0.31}$& 1328/994\\
00031403057 & 57198.15&--- &745 &   ---             &$  91.4 ^{\pm2.4} $ &        ---         &$  91.4^{\pm2.4} $  &$1.25^{\pm0.24}$&    ---         &    ---         &  ---           &$1.67^{\pm0.25}$&   39/50  \\
00031403058 & 57198.93&1312&--- &$  4592^{\pm12}$   &   ---              &$   1.34^{\pm0.11} $&$  4590^{\pm12}  $  &$0.95^{\pm0.09}$&$0.47^{\pm0.05}$&$0.86^{\pm0.09}$&$0.98^{\pm0.12}$&$1.70^{\pm0.35}$& 1322/928\\
00646721000 & 57199.52&--- &965 &   ---                &$  468.9 ^{\pm4.1}$ &        ---         &$  468.9^{\pm4.1}$  &$1.22^{\pm0.17}$&    ---         &    ---         &$0.58^{\pm0.08}$&$1.82^{\pm0.27}$ &  41/49  \\

\hline
\hline
\end{tabular}
\noindent{
\leftline{In Col. 2, UT dates of the year 2015 are mentioned in dd-mm format. In Cols. 4 \& 5, XRT and exposures are mentioned.}
\leftline{In Cols. 6 \& 7, XRT ($0.5-10$~keV) and BAT ($15-150$~keV) fluxes are mentioned. In Cols. 8 \& 9, model fitted disk 
blackbody (DBB) and powerlaw (PL)}
\leftline{fluxes are mentioned.}
\leftline{All fluxes are in the units of $10^{-9}$ $ergs~cm^{-2}~s^{-1}$}.
\leftline{${n_H}^1$ and ${n_H}^2$ are in the unit of $10^{22}$ $cm^{-2}$. ${n_H}^1$ is Hydrogen column density for 
interstellar absorption.${n_H}^2$ is Hydrogen column density for partial covering absorption.}
\leftline{The Fe emission  line energy and $\sigma$ are mentioned in Col. 15 \& 16.}
\leftline{Best fitted values of $\chi^2$ and degrees of freedom are mentioned in Col. 17 as $\chi^2/dof$.}
}
\end{table}
% \end{landscape}

% \begin{landscape}
\begin{table}
\vskip 0.5cm
\addtolength{\tabcolsep}{-4.50pt}
\scriptsize
{\large \bf Table II: TCAF Model Fitted Spectral Analysis Results}
\vskip 0.2cm
\centering
\begin{tabular}{lccccccccccccccc}
\hline
Obs ID& UT Date&Day&${n_{H}}^1$&${n_{H}}^2$&$CF$ &$\dot{m_d}$&$\dot{m_h}$&ARR&$X_{S}$& $R$ &$N$&Fe.Line&LW&$\chi^2 /dof$\\
  &(DD-MM) &($MJD$)&$(\times 10^{22})$ &$(\times 10^{22})$&  &($\dot{M}_{Edd}$) &($\dot{M}_{Edd}$)& ($\dot{m_h}$/$\dot{m_d}$) &($r_s$)  & &  & $(keV)$ & $(keV)$& \\
 (1)&  (2)  & (3)  & (4)& (5) & (6) & (7) & (8) & (9)& (10)& (11)&(12) &(13)& (14) & (15) \\
\hline

00643949000& 15-06 & 57188.77&$1.49^{\pm0.17}$&   ---         &   ---          &$ 0.05^{\pm0.01}$&$  0.85^{\pm0.04}$&$  17.0^{\pm1.0}$&$  334^{\pm15}$ &$ 2.56^{\pm0.10}$&$   15.5^{\pm1.2}$&    ---           &  ---             &    69/51  \\ 
00031403035& 16-06 & 57189.62&$1.86^{\pm0.22}$&$28.8^{\pm1.8}$&$0.41^{\pm0.03}$&$ 0.06^{\pm0.01}$&$  0.88^{\pm0.04}$&$  15.2^{\pm0.8}$&$  305^{\pm13}$ &$ 2.40^{\pm0.10}$&$   10.2^{\pm0.9}$&    ---           &  ---             &  708/899  \\
00031403038& 18-06 & 57191.01&$0.87^{\pm0.10}$&$ 2.2^{\pm0.1}$&$0.93^{\pm0.10}$&$ 0.06^{\pm0.01}$&$  0.90^{\pm0.03}$&$  15.3^{\pm0.4}$&$  275^{\pm10}$ &$ 2.60^{\pm0.12}$&$   17.6^{\pm0.8}$&$  6.57^{\pm0.11}$&$ 0.89^{\pm0.08}$ &  929/937  \\
00644520000& 18-06 & 57191.03&$1.27^{\pm0.14}$&$ 9.3^{\pm0.4}$&$0.63^{\pm0.07}$&$ 0.12^{\pm0.02}$&$  0.85^{\pm0.03}$&$   6.9^{\pm0.2}$&$  300^{\pm11}$ &$ 2.59^{\pm0.12}$&$   19.7^{\pm0.9}$&$  6.41^{\pm0.15}$&$ 1.10^{\pm0.17}$ &  338/281  \\
00031403042& 19-06 & 57192.16&$1.20^{\pm0.11}$&$12.2^{\pm0.8}$&$0.91^{\pm0.04}$&$ 0.10^{\pm0.01}$&$  1.37^{\pm0.07}$&$  13.4^{\pm0.7}$&$  302^{\pm12}$ &$ 1.91^{\pm0.08}$&$   24.0^{\pm1.1}$&$  6.44^{\pm0.15}$&$ 0.74^{\pm0.13}$ &  932/787  \\
00645176000& 20-06 & 57193.56&$0.58^{\pm0.05}$&   ---         &   ---          &$ 0.07^{\pm0.01}$&$  1.58^{\pm0.07}$&$  22.2^{\pm1.2}$&$  220^{\pm14}$ &$ 2.14^{\pm0.09}$&$   14.4^{\pm0.8}$&   ---            & ---              &    54/49  \\
00031403048& 21-06 & 57194.16&$0.57^{\pm0.09}$&   ---         &   ---          &$ 0.07^{\pm0.01}$&$  1.53^{\pm0.07}$&$  21.6^{\pm1.1}$&$  126^{\pm 9}$ &$ 1.92^{\pm0.11}$&$   12.4^{\pm0.4}$&$  6.46^{\pm0.29}$&$ 0.45^{\pm0.04}$ & 1086/899  \\
00031403046& 21-06 & 57194.54&$0.62^{\pm0.10}$&   ---         &   ---          &$ 0.08^{\pm0.01}$&$  1.58^{\pm0.08}$&$  20.2^{\pm1.2}$&$  141^{\pm 7}$ &$ 1.88^{\pm0.09}$&$   23.8^{\pm1.5}$&$  6.39^{\pm0.24}$&$ 1.20^{\pm0.22}$ & 1399/797  \\
00031403045& 22-06 & 57195.08&$0.55^{\pm0.07}$&$ 6.2^{\pm0.2}$&$0.86^{\pm0.09}$&$ 0.07^{\pm0.01}$&$  1.54^{\pm0.05}$&$  21.1^{\pm0.9}$&$  162^{\pm10}$ &$ 2.18^{\pm0.11}$&$   18.3^{\pm0.6}$&$  6.92^{\pm0.22}$&$ 1.40^{\pm0.15}$ & 1117/928  \\
00031403049& 22-06 & 57195.35&$0.66^{\pm0.06}$&$18.0^{\pm0.3}$&$0.66^{\pm0.05}$&$ 0.08^{\pm0.01}$&$  1.57^{\pm0.05}$&$  20.3^{\pm0.7}$&$  217^{\pm12}$ &$ 3.01^{\pm0.09}$&$   12.3^{\pm0.5}$&$  6.91^{\pm0.18}$&$ 1.19^{\pm0.20}$ & 1067/707  \\
00031403047& 22-06 & 57195.41&$0.63^{\pm0.05}$&   ---         &   ---          &$ 0.09^{\pm0.01}$&$  1.24^{\pm0.05}$&$  13.6^{\pm0.6}$&$  269^{\pm14}$ &$ 2.68^{\pm0.12}$&$   18.9^{\pm0.9}$&$  6.97^{\pm0.23}$&$ 0.86^{\pm0.11}$ &  982/709  \\
00031403052& 23-06 & 57196.89&$0.84^{\pm0.10}$&$ 5.6^{\pm0.5}$&$0.95^{\pm0.15}$&$ 0.10^{\pm0.01}$&$  1.26^{\pm0.03}$&$  12.7^{\pm0.3}$&$  290^{\pm11}$ &$ 3.16^{\pm0.11}$&$   22.1^{\pm0.9}$&$  6.45^{\pm0.21}$&$ 0.30^{\pm0.05}$ & 1001/937  \\
00031403054& 24-06 & 57197.21&$1.10^{\pm0.14}$&$28.9^{\pm1.3}$&$0.73^{\pm0.07}$&$ 0.08^{\pm0.01}$&$  1.54^{\pm0.04}$&$  19.5^{\pm0.9}$&$  187^{\pm 9}$ &$ 2.88^{\pm0.10}$&$   26.5^{\pm1.2}$&$  6.64^{\pm0.20}$&$ 0.61^{\pm0.14}$ & 1405/981  \\
00031403053& 24-06 & 57197.33&$0.54^{\pm0.04}$&   ---         &   ---          &$ 0.11^{\pm0.01}$&$  1.44^{\pm0.07}$&$  13.4^{\pm0.6}$&$  168^{\pm 7}$ &$ 2.63^{\pm0.10}$&$   20.3^{\pm1.5}$&$  6.55^{\pm0.18}$&$ 0.29^{\pm0.06}$ & 1213/991  \\
00031403055& 25-06 & 57198.00&$0.85^{\pm0.07}$&$ 5.2^{\pm0.4}$&$0.85^{\pm0.06}$&$ 0.15^{\pm0.01}$&$  1.39^{\pm0.05}$&$   9.0^{\pm0.4}$&$  132^{\pm 8}$ &$ 2.22^{\pm0.09}$&$   37.5^{\pm2.7}$&$  6.08^{\pm0.14}$&$ 0.96^{\pm0.13}$ &  937/727  \\
00031403056& 25-06 & 57198.02&$0.60^{\pm0.04}$&$ 8.2^{\pm0.8}$&$0.81^{\pm0.06}$&$ 0.15^{\pm0.01}$&$  1.36^{\pm0.04}$&$   8.9^{\pm0.3}$&$  154^{\pm 7}$ &$ 2.04^{\pm0.12}$&$   59.6^{\pm3.6}$&$  6.30^{\pm0.19}$&$ 1.11^{\pm0.16}$ & 1303/994  \\
00031403057& 25-06 & 57198.15&$1.20^{\pm0.11}$&   ---         &   ---          &$ 0.15^{\pm0.01}$&$  1.57^{\pm0.06}$&$  10.6^{\pm0.4}$&$  174^{\pm10}$ &$ 2.08^{\pm0.07}$&$   93.4^{\pm2.9}$&    ---           &   ---            &    41/51  \\
00031403058& 25-06 & 57198.93&$0.87^{\pm0.04}$&$ 3.9^{\pm0.5}$&$0.92^{\pm0.05}$&$ 0.15^{\pm0.01}$&$  1.45^{\pm0.05}$&$  10.2^{\pm0.4}$&$  160^{\pm 9}$ &$ 1.88^{\pm0.10}$&$   89.5^{\pm4.5}$&$  6.22^{\pm0.22}$&$ 0.90^{\pm0.21}$ & 1423/943  \\
00646721000& 26-06 & 57199.52&$1.15^{\pm0.16}$&   ---         &   ---          &$ 0.13^{\pm0.02}$&$  1.31^{\pm0.05}$&$   9.8^{\pm0.4}$&$  143^{\pm 8}$ &$ 1.99^{\pm0.09}$&$   77.5^{\pm3.3}$&    ---           &    ---           &   36/51   \\

\hline

\hline
\end{tabular}
\noindent{
\leftline{In Col. 2, UT dates of the year 2015 are mentioned in dd-mm format.}
\leftline{${n_H}^1$ and ${n_H}^2$ are in the unit of $10^{22}$ $cm^{-2}$. ${n_H}^1$ is Hydrogen column density for 
interstellar absorption.${n_H}^2$ is Hydrogen column density for partial covering absorption.}
\leftline{TCAF model fitted/derived parameters are mentioned in Cols. 7-12. The Fe emission  line energy and
$\sigma$ are mentioned in Col. 13 \& 14.}
\leftline{Best fitted values of $\chi^2$ and degrees of freedom are mentioned in Col. 15 as $\chi^2/dof$.}
\leftline{Note: Mass of the black hole was kept frozen at $10.6$ $M_{\odot}$ during spectral fitting with
the TCAF model fits file.}
\leftline{The average values of 90\% confidence $\pm $ values obtained using {\tt steppar} command in XSPEC, are placed
as superscripts of fitted parameter values.}
}
\end{table}
% \end{landscape}

\clearpage

\begin{table}
\vskip -0.0cm
\addtolength{\tabcolsep}{-4.50pt}
\scriptsize
\centering
\centering{\large \bf Table III: Time Resolved BAT Spectra}
\vskip 0.2cm
\begin{tabular}{lcccccccccccc}
\hline
Obs ID& Day&Spectra& Exposures&$\dot{m_d}$&$\dot{m_h}$&ARR&$X_{S}$& $R$ &$N$&$\chi^2 /dof$\\
  & ($MJD$)& & (sec) &  ($\dot{M}_{Edd}$) &($\dot{M}_{Edd}$)& & ($r_s$)&  & & & \\
 (1)&  (2)  & (3)  & (4)& (5) & (6) & (7) & (8) & (9)& (10)& (11) \\
\hline
00643949000&  57188.77&   & 1202 &$  0.050^{\pm0.0013}$&$ 0.852^{\pm0.046}$&$  17.04^{\pm0.98}$&$ 334^{\pm15}$&$ 2.55^{\pm0.10}$&$ 15.49^{\pm1.19}$&  69/51\\ 
           &          &  1&  220 &$  0.055^{\pm0.0015}$&$ 0.879^{\pm0.043}$&$  15.98^{\pm0.85}$&$ 326^{\pm13}$&$ 2.63^{\pm0.09}$&$ 15.58^{\pm1.22}$&  66/51\\
           &          &  2&  680 &$  0.057^{\pm0.0014}$&$ 0.823^{\pm0.045}$&$  14.44^{\pm0.71}$&$ 336^{\pm13}$&$ 2.37^{\pm0.10}$&$ 12.55^{\pm1.09}$&  72/51\\
           &          &  3&  160 &$  0.041^{\pm0.0015}$&$ 0.876^{\pm0.043}$&$  21.36^{\pm0.91}$&$ 330^{\pm13}$&$ 2.55^{\pm0.09}$&$ 19.76^{\pm1.51}$&  59/51\\
\hline
00644520000&  57191.03&   & 1202 &$  0.124^{\pm0.0022}$&$ 0.851^{\pm0.028}$&$   6.92^{\pm0.22}$&$ 300^{\pm11}$&$ 2.59^{\pm0.10}$&$ 19.69^{\pm0.89}$&  338/281\\
           &          &  1&  200 &$  0.131^{\pm0.0022}$&$ 0.854^{\pm0.037}$&$    6.52^{\pm0.39}$&$ 317^{\pm10}$&$ 2.81^{\pm0.11}$&$ 21.54^{\pm1.17}$&  56/51\\
           &          &  2&  220 &$  0.109^{\pm0.0026}$&$ 0.847^{\pm0.053}$&$    7.77^{\pm0.45}$&$ 269^{\pm 10}$&$ 2.43^{\pm0.12}$&$ 15.76^{\pm1.34}$&  61/51\\
           &          &  3&   50 &$  0.096^{\pm0.0025}$&$ 0.909^{\pm0.049}$&$   9.46^{\pm0.51}$&$ 322^{\pm11}$&$ 2.50^{\pm0.13}$&$ 22.42^{\pm1.81}$&  71/51\\
           &          &  4&  150 &$  0.124^{\pm0.0037}$&$ 0.822^{\pm0.045}$&$    6.63^{\pm0.42}$&$ 309^{\pm11}$&$ 2.41^{\pm0.11}$&$ 20.36^{\pm1.15}$&  57/51\\
\hline
00645176000&  57193.56&   &  542 &$  0.071^{\pm0.0037}$&$ 1.577^{\pm0.068}$&$  22.15^{\pm1.17}$&$ 220^{\pm14}$&$ 2.14^{\pm0.09}$&$ 14.43^{\pm0.77}$&  54/49 \\
           &          &  1&  150 &$  0.071^{\pm0.0044}$&$ 1.582^{\pm0.068}$&$  22.28^{\pm1.19}$&$ 218^{\pm14}$&$ 2.15^{\pm0.09}$&$ 14.04^{\pm0.76}$&  62/49 \\
           &          &  2&  260 &$  0.071^{\pm0.0036}$&$ 1.553^{\pm0.075}$&$  21.87^{\pm1.22}$&$ 230^{\pm15}$&$ 2.13^{\pm0.09}$&$ 14.93^{\pm0.90}$&  65/49 \\
\hline
00646721000&  57199.52&   &  965 &$  0.134^{\pm0.0207}$&$ 1.311^{\pm0.047}$&$   9.79^{\pm0.17}$&$ 143^{\pm 8}$&$ 1.99^{\pm0.09}$&$ 77.55^{\pm3.29}$&  35/51\\
           &          &  1&  530 &$  0.136^{\pm0.0212}$&$ 1.312^{\pm0.051}$&$   9.65^{\pm0.22}$&$ 142^{\pm 8}$&$ 2.01^{\pm0.10}$&$ 74.15^{\pm3.37}$&  37/51\\
           &          &  2&   40 &$  0.133^{\pm0.0194}$&$ 1.310^{\pm0.042}$&$   9.85^{\pm0.18}$&$ 144^{\pm 9}$&$ 1.96^{\pm0.09}$&$ 89.22^{\pm2.54}$&  36/51\\
\hline

\end{tabular}
\noindent{
\leftline{TCAF fitted extracted parameters for time resolved BAT spectra in the energy range of $15-150$~keV.} 
\leftline{Note: First row in each spectra are TCAF model fitted spectral analysis results when entire data exposure including gaps are used.} 
\leftline{The mass of the BH is frozen at $10.6$ $M_{\odot}$ during the fitting.}
}
\end{table}

\begin{table}
\vskip 0.5cm
\addtolength{\tabcolsep}{-4.50pt}
\scriptsize
\centering
\centering{\large \bf Table IV}
\vskip 0.2cm
\begin{tabular}{lcccccccccccc}
\hline
Obs ID& Day&$n_e$& $B$ &$\Lambda_{syn} $&$t_{sync}$&$\Lambda_{Comp} $&$t_{Comp} $&$t_{inf}$ & $t_{sync}/t_{inf}$&$t_{Comp}/t_{inf}$ \\
  & ($MJD$)&$(\times 10^{16})$ &($Gauss$) &$ (\times 10^{-9} ergs/sec)$&($s$)&$ (\times 10^{-3} ergs/s)$&$(\times 10^{-3} s)$&($s$)& & \\
 (1)&  (2)  & (3)  & (4)& (5) & (6) & (7) & (8) & (9) & (10) & (11)\\
\hline
00643949000&57188.77 &$ 1.07$&$  97^{\pm3}$&$1.01 $&$  81.16$&$0.25 $&$12.34$&$  1.22$ &$   66.48$&$10.10$\\ 
00031403035&57189.62 &$ 1.51$&$ 124^{\pm4}$&$1.62 $&$  50.51$&$0.39 $&$9.06 $&$  1.04$ &$   48.44$&$8.69 $\\
00031403038&57191.01 &$ 2.05$&$ 149^{\pm4}$&$2.35 $&$  34.91$&$0.45 $&$8.38 $&$  0.91$ &$   38.33$&$9.20 $\\
00644520000&57191.03 &$ 1.59$&$ 125^{\pm4}$&$1.66 $&$  49.24$&$0.69 $&$4.97 $&$  1.04$ &$   47.30$&$4.77 $\\
00031403042&57192.16 &$ 2.28$&$ 153^{\pm5}$&$2.50 $&$  32.79$&$0.53 $&$6.74 $&$  1.05$ &$   31.24$&$6.42 $\\
00645176000&57193.56 &$ 6.67$&$ 308^{\pm12}$&$10.12$&$   8.17$&$0.67 $&$7.34 $&$  0.65$ &$   12.53$&$11.20$\\ 
00031403048&57194.16 &$34.38$&$ 923^{\pm44}$&$90.14$&$   0.91$&$3.07 $&$2.79 $&$  0.28$ &$    3.21$&$9.85 $\\
00031403046&57194.54 &$25.57$&$ 752^{\pm25}$&$59.91$&$   1.37$&$2.52 $&$3.04 $&$  0.33$ &$    4.09$&$9.10 $\\
00031403045&57195.08 &$16.55$&$ 564^{\pm19}$&$33.71$&$   2.43$&$1.49 $&$4.46 $&$  0.41$ &$    5.92$&$10.91$\\
00031403049&57195.35 &$ 7.28$&$ 305^{\pm9}$&$9.82 $&$   8.34$&$0.53 $&$8.38 $&$  0.64$ &$   12.99$&$13.04$\\ 
00031403047&57195.41 &$ 3.03$&$ 182^{\pm7}$&$3.49 $&$  23.46$&$0.50 $&$7.50 $&$  0.88$ &$   26.59$&$8.49 $\\
00031403052&57196.89 &$ 2.56$&$ 154^{\pm5}$&$4.21 $&$  19.49$&$1.67 $&$9.49 $&$  0.99$ &$   19.71$&$1.95 $\\
00031403054&57197.21 &$11.29$&$ 414^{\pm14}$&$18.22$&$   4.51$&$0.84 $&$6.25 $&$  0.51$ &$    8.85$&$12.33$\\ 
00031403053&57197.33 &$14.49$&$ 504^{\pm16}$&$26.91$&$   3.04$&$1.66 $&$3.65 $&$  0.44$ &$    6.99$&$8.83 $\\
00031403055&57198.00 &$28.70$&$ 820^{\pm29}$&$71.15$&$   1.51$&$5.06 $&$1.60 $&$  0.31$ &$    3.78$&$5.26 $\\
00031403056&57198.02 &$17.84$&$ 602^{\pm24}$&$38.33$&$   2.13$&$3.78 $&$1.86 $&$  0.38$ &$    5.59$&$4.86 $\\
00031403057&57198.15 &$13.94$&$ 500^{\pm18}$&$26.41$&$   3.09$&$2.35 $&$2.63 $&$  0.46$ &$    6.72$&$5.72 $\\ 
00031403058&57198.93 &$16.65$&$ 568^{\pm26}$&$34.21$&$   2.40$&$3.31 $&$2.03 $&$  0.41$ &$    5.89$&$4.99 $\\ 
00646721000&57199.55 &$21.38$&$ 685^{\pm25}$&$49.64$&$   1.65$&$4.31 $&$1.76 $&$  0.34$ &$    4.84$&$5.16 $\\
\hline
\end{tabular}
\noindent{
\leftline {Electron number density ($n_e$) is in the unit of $\times 10^{16}$ cm$^3$.}
}
\end{table}

\label{lastpage}


\begin{thebibliography}{99}
%% you can type \apj for ApJ, \aap for A&A, \apss for Ap&SS, etc. Please consult
%% the macro chjaa.cls. You can also find them in aasguide.tex (AASTeX for ApJ, AJ, PASP)
%% Please follow the format of ChJAA's reference list

\bibitem[Arnaud 1996]{Arnaud96} Arnaud, K.A., 1996, ASP Conf. Ser., Astronomical Data Analysis Software and Systems V, ed. G.H. Jacoby \& J. Barnes, 101, 17

\bibitem[Barthelmy  et al. 2015]{Barthelmy2015}Barthelmy S. D., D’Ai A., D’Avanzo P., Krimm H. A., Lien A. Y., Marshall F. E., Maselli A., Siegel M. H., 2015, GCN Circ., 17929

\bibitem[Beardmore et al. 2015]{Beardmore et al. 2015} Beardmore A. P., Altamirano D., \& Kuulkers E., et al. 2015, ATel, 736, 1

\bibitem[Bhattacharjee et al. 2017]{Bhattacharjee et al. 2017} Bhattacharjee, A., Banerjee, I., \& Banerjee, A., et al. 2017, MNRAS, 466, 1372

\bibitem[Casares et al. 1992]{Casaries1992} Casares J., Charles P. A., Naylor T., 1992, Nature, 355, 614

\bibitem[Chakrabarti 1989]{C89} Chakrabarti, S. K., 1989, MNRAS, 240, 7

\bibitem[Chakrabarti 1990]{C90} Chakrabarti, S. K. 1990, Theory of Transonic Astrophysical Flows (Singapore: World Scientific)

\bibitem[Chakrabrti \& Molteni 1993]{CM93} Chakrabarti, S. K.; Molteni, D., 1993, ApJ, 417, 671

\bibitem[Chakrabarti \& D'Silva 1994]{CD94} Chakrabarti, S. K., \& D'Silva, S. 1994, ApJ, 424, 138

\bibitem[Chakrabarti 1995]{C95}Chakrabarti, S.K., 1995, in Ann. NY Acad. Sci., Seventeenth Texas Symposium on Relativistic Astrophysics and 
Cosmology, eds. H. Bohringer, G.E. Morfil and J. Trumper, 546.

\bibitem[Chakrabarti \& Titarchuk 1995]{CT95} Chakrabarti, S. K., \& Titarchuk, L.G., 1995, ApJ, 455, 623

\bibitem[Chakrabarti 1997]{C97} Chakrabarti, S.K., 1997, ApJ, 484, 313

\bibitem[Chakrabarti 1999]{C99} Chakrabarti, S. K., 1999, A\&A, 351, 185 

\bibitem[Chakrabarti et al. 2015]{C15} Chakrabarti, S. K., Mondal, S., \& Debnath, D., 2015, MNRAS, 452, 3451

\bibitem[Chakrabarti 2018]{C18}Chakrabarti, S. K., 2018, MG14 Conf. Pro., World Scientific Press, Singapore, Edited by, R. Ruffini, R. Jantzen, M. Bianchi, 369

\bibitem[Chakrabarti et al. 2019]{Chakrabarti19} Chakrabarti, S. K., Nagarkoti, S., \& Debnath, D., 2019, AdSpR, 63, 3749

\bibitem[Chatterjee et al. 2016]{C16} Chatterjee, D., Debnath, D., Chakrabarti, S. K., et al. 2016, ApJ, 827, 88

\bibitem[Chatterjee et al. 2019]{C19} Chatterjee, D., Debnath, D., Jana, A., \& Chakrabarti, S. K., 2019, Ap\&SS, 364, 14

\bibitem[Chatterjee et al. 2020]{C20} Chatterjee, K., Debnath, D., \& Chatterjee, D., et al., 2020, MNRAS, 493, 2452

\bibitem[Chatty et al. 2011]{Chatty11} Chaty, S., Dubus, G., \& Raichoor, A., 2011, A\&A, 529, 3

\bibitem[Cutri et al. 2003]{Cutri03} Cutri, R. M., Skrutskie, M. F., \&  van Dyk, S., et al. 2003, VizieR Online Data Catalog. no. 2246

\bibitem[Dallilar et al. 2017]{Dallilar17} Dallilar, Y., Eikenberry, S. S., \& Garner, A., et al. 2017, Sci, 358, 1299

\bibitem[Debnath et al. 2014]{DCM14} Debnath, D., Mondal, S., \& Chakrabarti, S. K., 2014, MNRAS, 440, L121

\bibitem[Debnath et al. 2015a]{DMC15} Debnath, D., Mondal, S., \& Chakrabarti, S. K., 2015a, MNRAS, 447, 1984

\bibitem[Debnath et al. 2015b]{DMCM15} Debnath, D., Molla, A.A., Chakrabarti, S.K., \& Mondal, S., 2015b, ApJ, 803, 59

\bibitem[Debnath et al. 2017]{D17} Debnath, D., Jana, A., Chakrabarti, S. K., \& Chatterjee, D., 2017, ApJ, 850, 92

\bibitem[Debnath et al. 2020]{D20} Debnath, D., Chatterjee, D. Jana, A., Chakrabarti, S. K., \& Chatterjee, K., 2020, RAA (in press)

\bibitem[Del Santo et al. 2013]{Del Santo13} Del Santo, M., Malzac, J., \& Belmont, R., et al. MNRAS, 430, 209

\bibitem[D'Silva \& Chakrabarti 1994]{DC94} D'Silva, S., Chakrabarti, S. K., 1994, ApJ, 424, 149

\bibitem[Esin et al. 1997]{Esin97} Esin, A. A., McClintock, J. E., Narayan, R., 1997, ApJ, 489, 865

\bibitem[Gandhi et al. 2016]{Gandhi16} Gandhi, P., Littlefair, S. P., \& Hardy, L. K., et al. 2016, MNRAS, 459, 554

\bibitem[Garain et al. 2012]{Garain12} Garain, S. K., Ghosh, H., \& Chakrabarti, S. K., 2012, ApJ, 758, 114

\bibitem[Gazeas et al. 2015]{Gazeas15} Gazeas K., Vasilopoulos G., Petropoulou M., \& Sapountzis K., 2015, ATel, 7650, 1

\bibitem[Giri et al. 2010]{Giri10} Giri, K., Chakrabarti, S. K., Samanta, M. M., Ryu, D., 2010, MNRAS, 403, 516

\bibitem[Haardt \& Marschi 1993]{HM93} Haardt, F., \& Maraschi, L., 1993, ApJ, 413, 507

\bibitem[Heinz1 et al. 2016]{Heinz16} Heinz, S., Corrales, L., Smith, R., et al. 2016, ApJ, 825, 15

\bibitem[Huppenkothen et al. 2017]{Huppenkothen17} Huppenkothen, D., Younes, G., Ingram, A., et al. 2017, ApJ, 834, 90

\bibitem[Jana et al. 2016]{Jana16} Jana, A., Debnath, D., Chakrabarti, S. K., et al. 2016, ApJ, 819, 107 

\bibitem[Jana et al. 2017]{JCD17} Jana, A., Chakrabarti, S. K., \& Debnath, D., 2017, ApJ, 850, 91 (JCD17)

\bibitem[Jana et al. 2020a]{Jana20a} Jana, A., Debnath, D., Chakrabarti, S. K., \& Chatterjee, D., 2020a, RAA, 20, 28

\bibitem[Jana et al. 2020b]{Jana20b} Jana, A., Debnath, D., \& Chatterjee, D., et al. 2020b, ApJ (in press)

\bibitem[Jenke1 et al. 2016]{Jenke16} Jenke P. A., Wilson-Hodge, C. A., \& Homan, J., et al., 2016, ApJ, 826, 37

\bibitem[King et al. 2015]{King15} King A. L., Miller J. M., \& Raymond J., et al. 2015, ApJ, 813, L37

\bibitem[Khargharia et al. 2010]{Khargharia10} Khargharia J., Froning C. S., Robinson E. L., 2010, ApJ, 716, 1105

\bibitem[Lipunov et al. 2015]{Lipunov15} Lipunov, V., Gorbovskoy, E., \& Tiurina, N., et al. 2015, Atel, 8453, 1

\bibitem[Loh et al. 2016]{Loh16} Loh, A., Corbel, S., \&  Dubus, G., et al. 2016, MNRAS, 462, L111

\bibitem[Makino 1989]{Makino89} Makino F., 1989, IAU Circ., 4782, 1

\bibitem[Miller-Jones et al. 2009]{Miller-Jones09} Miller-Jones J. C. A., Jonker P. G., \& Dhawan V., et al. 2009, ApJ, 706, L230

% \bibitem[Molla et al. (2017)]{Molla17} Molla, A. A., Debnath, D., \& Chakrabarti, S. K. et al. 2017, ApJ, 834, 88
\bibitem[Molla et al. 2016]{Molla16} Molla, A. A., Debnath, D., \& Chakrabarti, S. K. et al. 2016, MNRAS, 460. 3163

\bibitem[Mondal et al. 2014]{Mondal14} Mondal, S., Debnath, D., \& Chakrabarti, S.K., 2014, ApJ, 786, 4

\bibitem[Mondal et al. 2016]{Mondal16} Mondal, S., Chakrabarti, S.K., \& Debnath, D., 2016, Ap$\&$SS, 361, 309

\bibitem[Molteni et al. 2016]{MSC96} Molteni D., Sponholz H., \& Chakrabarti S. K., 1996, ApJ, 457, 805

\bibitem[Mooley et al. 2015a]{Mooley15a} Mooley K., Fender R., \& Anderson G., et al. 2015a, ATel, 7658, 1

\bibitem[Mooley et al. 2015b]{Mooley15b)} Mooley K., Clarke F., Fender R., 2015b, ATel, 7714, 1

\bibitem[Motta et al. 2017]{Motta17} Motta, S. E., Kajava, J. J. E., \& Sanchez-Fernandez, C., et al. 2017, MNRAS, 471, 1797

\bibitem[Nandi et al. 2012]{Nandi12} Nandi, A., Debnath, D., Mandal, S., et al.\ 2012, \aap, 542, 56

\bibitem[Narayan \& Yi 1994]{Narayan1994} Narayan, R., Yi, I., 1994, ApJ, 428, L13

\bibitem[Natalucci et al. 2015]{Natalucci15}Natalucci L., Fiocchi M., Bazzano A., Ubertini P., Roques J.-P., Jourdain E., 2015, ApJL, 813, 
L21

\bibitem[Negoro et al. 2015]{Negoro15} Negoro, H., Matsumitsu, T., Mihara, T., et al.\ 2015, The Astronomer's Telegram 7646, 1

\bibitem[Novikov \& Thorne 1973]{NV73} Novikov, \& Thorne, 1973, Black Holes, (Eds.) C. DeWitt \& B. DeWitt (Gordon \& Breach: New York)

\bibitem[Radhika et al. 2016]{Radhika16} Radhika, D., Nandi, A., Agrawal, V. K., \& Mandal, S., 2016, MNRAS, 462, 1834

\bibitem[Remilard \& McClintock 2006]{RM06} Remillard R. A., McClintock J. E., 2006, ARA\&A, 44, 49

\bibitem[Ritcher 1989]{Ritcher89} Ritcher, G. A., 1989, 

\bibitem[Rodriguez et al. 2015]{Rodriguez15} Rodriguez, J., Cadolle Bel, M., \& Alfonso-Garzón, J., et al. 2015, A\&A, 581L, 9

\bibitem[Roques et al. 2015]{Roques15} Roques J.-P., Jourdain E., Bazzano A., Fiocchi M., Natalucci L., Ubertini P., 2015, ApJL, 813, L22

\bibitem[Ryu et al. 1997]{Ryu97} Ryu, D., Chakrabarti, S. K., \& Molteni, D. 1997, ApJ, 474, 378

\bibitem[Rybicki \& Lightman 1979]{Rybicki79} Rybicki, G. B., \& Lightman, A. P., 1979, Radiative processes in astrophysics (Wiley, New 
York)

\bibitem[Shahbaz et al. 1994]{Shahbaz94} Shahbaz T., Ringwald F. A., \& Bunn J. C., et al. 1994, MNRAS, 271, L10

\bibitem[Shakura \& Sunyaev 1973]{SS73} Shakura, N.I., \& Sunyaev, R. A., 1973, A\&A, 24, 337

\bibitem[Shang et al. 2019]{Shang19} Shang, J.R., Debnath, D., \& Chatterjee, D., et al., 2019, ApJ, 875, 4

\bibitem[Siegert et al. 2016]{Siegert16} Siegert, T., Diehl, R., \& Greiner, J., et al. 2016, Nature, 531, 341

\bibitem[Sivakoff et al. 2015]{Sivakoff15} Sivakoff G., Bahramian A., \& Altamirano D., et al. 2015, ATel, 7763, 1

\bibitem[Sunyaev \& Titarchuk 1980]{ST80} Sunyaev, R.A., \& Titarchuk, L. G., 1980, ApJ, 86, 121

\bibitem[Sunyaev \& Titarchuk 1985]{ST85} Sunyaev, R.A., \& Titarchuk, L. G., 1985, A\&A, 143, 374

\bibitem[Tetarenko et al. 2015]{Tetarenko15} Tetarenko A., Sivakoff G. R., \& Young K., et al. 2015, ATel, 7708, 1

\bibitem[Tetarenko et al. 2017]{Tetarenko17} Tetarenko, A. J., Sivakoff, G. R., Miller-Jones, J. C. A., et al. 2017, MNRAS, 469, 3141 

%\bibitem[Tingay et al. (1995)]{Tingay95}Tingay, S. J., et al. 1995, Nature, 374, 141
\bibitem[Trushkin et al. 2015a]{Trushkin15a} Trushkin S. A., Nizhelskij N. A., \& Tsybulev P. G., 2015a, ATel, 7667, 1

\bibitem[Trushkin et al. 2015b]{Trushkin15b} Trushkin S. A., Nizhelskij N. A., \& Tsybulev P. G., 2015b, ATel, 7716, 1

\bibitem[Tsubono et al. 2015]{Tsubono15} Tsubono K., Aoki T., \& Asuma K., et al. 2015, ATel, 7701, 1

\bibitem[Vasilopoulos et al. 2016]{Vasilopoulos16} Vasilopoulos G., \& Petropoulou M., 2016, MNRAS, 455, 4426

\bibitem[Wachmann 1948]{Wachmann48} Wachmann, A. A., 1948, Erg. Astron. Nachr., 11, 5

\bibitem[Walton et al. 2017]{Walton17} Walton, D. J., Mooley, K., \& King, A. L., et al. 2017, 839, 110
%\bibitem[Yang (2012a)]{Yang12a} Yang, Y. J., Wijnands, R., Kennea, J. A., 2012a, ATel, 3975, 1
%\bibitem[Yang (2012b)]{Yang12b} Yang, J., Xu, Y., Li, Z., 2012b, MNRAS, 426, 66
%\bibitem[Znajek (1978)]{Z78} Znajek, R. L., 1978, MNRAS, 182, 639

\bibitem[Zdziarski 1993]{Z93} Zdziarski, A. A., Zycki, Piotr T., Svensson, R., Boldt, E., 1993, ApJ, 405, 125

\end{thebibliography}
\end{document}